\documentclass[12pt]{article}

\usepackage{amsmath, amssymb, slashed, epsf, color, graphicx, subfigure}

\makeatletter
\renewcommand{\section}{%
  \@startsection{section}%
   {1}%
   {\z@}%
   {-2.5ex \@plus -1ex \@minus -.2ex}%
   {2.3ex \@plus.2ex}%
   {\normalfont\large\bfseries}%
}%
\makeatother

\makeatletter
\renewcommand{\subsection}{%
  \@startsection{subsection}%
   {1}%
   {\z@}%
   {-2.5ex \@plus -1ex \@minus -.2ex}%
   {2.3ex \@plus.2ex}%
   {\normalfont\normalsize\bfseries}%
}%
\makeatother

\begin{document}

%%%%%%%%%%%%%%%%%%%%%%%%%%%%%%%%%%%%%%%%%%%

\def\a{\alpha}
\def\b{\beta}
\def\c{\varepsilon}
\def\d{\delta}
\def\e{\epsilon}
\def\f{\phi}
\def\g{\gamma}
\def\h{\theta}
\def\k{\kappa}
\def\l{\lambda}
\def\m{\mu}
\def\n{\nu}
\def\p{\psi}
\def\q{\partial}
\def\r{\rho}
\def\s{\sigma}
\def\t{\tau}
\def\u{\upsilon}
\def\v{\varphi}
\def\w{\omega}
\def\x{\xi}
\def\y{\eta}
\def\z{\zeta}
\def\D{\Delta}
\def\G{\Gamma}
\def\H{\Theta}
\def\L{\Lambda}
\def\F{\Phi}
\def\P{\Psi}
\def\S{\Sigma}

\def\o{\over}
\def\beq{\begin{eqnarray}}
\def\eeq{\end{eqnarray}}
\newcommand{\gsim}{ \mathop{}_{\textstyle \sim}^{\textstyle >} }
\newcommand{\lsim}{ \mathop{}_{\textstyle \sim}^{\textstyle <} }
\newcommand{\vev}[1]{ \left\langle {#1} \right\rangle }
\newcommand{\bra}[1]{ \langle {#1} | }
\newcommand{\ket}[1]{ | {#1} \rangle }
\newcommand{\EV}{ {\rm eV} }
\newcommand{\KEV}{ {\rm keV} }
\newcommand{\MEV}{ {\rm MeV} }
\newcommand{\GEV}{ {\rm GeV} }
\newcommand{\TEV}{ {\rm TeV} }
\newcommand{\1}{\mbox{1}\hspace{-0.25em}\mbox{l}}
\def\diag{\mathop{\rm diag}\nolimits}
\def\Spin{\mathop{\rm Spin}}
\def\SO{\mathop{\rm SO}}
\def\O{\mathop{\rm O}}
\def\SU{\mathop{\rm SU}}
\def\U{\mathop{\rm U}}
\def\Sp{\mathop{\rm Sp}}
\def\SL{\mathop{\rm SL}}
\def\tr{\mathop{\rm tr}}

\def\IJMP{Int.~J.~Mod.~Phys. }
\def\MPL{Mod.~Phys.~Lett. }
\def\NP{Nucl.~Phys. }
\def\PL{Phys.~Lett. }
\def\PR{Phys.~Rev. }
\def\PRL{Phys.~Rev.~Lett. }
\def\PTP{Prog.~Theor.~Phys. }
\def\ZP{Z.~Phys. }

\def\dd{\mathrm{d}}
\def\ff{\mathrm{f}}
\def\BH{{\rm BH}}
\def\inf{{\rm inf}}
\def\ev{{\rm evap}}
\def\eq{{\rm eq}}
\def\SM{{\rm sm}}
\def\Mpl{M_{\rm Pl}}
\def\GeV{{\rm GeV}}
\newcommand{\Red}[1]{\textcolor{red}{#1}}

\def\mDM{m_{\rm DM}}
\def\mphi{m_{\phi}}
\def\TeV{{\rm TeV}}
\def\Gphi{\Gamma_\phi}
\def\TR{T_{\rm RH}}
\def\Br{{\rm Br}}
\def\DM{{\rm DM}}
\def\Eth{E_{\rm th}}
\newcommand{\lmk}{\left(}  
\newcommand{\rmk}{\right)}
\newcommand{\lkk}{\left[}  
\newcommand{\rkk}{\right]}
\newcommand{\lhk}{\left \{ }  
\newcommand{\rhk}{\right \} }
\newcommand{\del}{\partial}  
\newcommand{\la}{\left\langle} 
\newcommand{\ra}{\right\rangle}

%% DIRAC SLASH %%%%%%%%%%%%%%%%%%%
\def\slashchar#1{\setbox0=\hbox{$#1$} % set a box for #1
\dimen0=\wd0 % and get its size
\setbox1=\hbox{/} \dimen1=\wd1 % get size of /
\ifdim\dimen0>\dimen1 % #1 is bigger
\rlap{\hbox to \dimen0{\hfil/\hfil}} % so center / in box
#1 % and print #1
\else % / is bigger
\rlap{\hbox to \dimen1{\hfil$#1$\hfil}} % so center #1
/ % and print /
\fi}

%%%%%%%%%%%%%%%%%%%%%%%%%%%%%%%%%%%%%%%%%%%%%%%%%%%%%%%%%%%%%%%

\begin{titlepage}
\begin{center}

\hfill IPMU-14-0343 \\
\hfill ICRR-Report-695-2014-21\\
\hfill \today

\vspace{1.5cm}
{\large\bf 
Coupling Unification and Dark Matter in a Standard Model Extension 
with Adjoint Majorana Fermions 
}

\vspace{2.0cm}
{\bf Tasuku Aizawa}$^{(a)}$,
{\bf Masahiro Ibe}$^{(a, b)}$,
and
{\bf Kunio Kaneta}$^{(a)}$

\vspace{2.0cm}
{\it
$^{(a)}${Institute for Cosmic Ray Research (ICRR), Theory Group, \\
University of Tokyo, Kashiwa, Chiba 277-8568, Japan} \\
$^{(b)}${Kavli Institute for the Physics and Mathematics of the Universe (IPMU),\\
University of Tokyo, Kashiwa, Chiba 277-8583, Japan} \\
}

\vspace{1.5cm}
\abstract{
We revisit an extension of the Standard Model with Majorana fermions in the adjoint representations.
There, a precise coupling unification and the good candidate 
for dark matter (the $SU(2)_L$ triplet fermion) are achieved simultaneously.
In particular, we show that the $SU(3)_c$ octet fermion which is required for  successful unification 
can be a good non-thermal source of the triplet fermion dark matter.
We also show that the scenario predicts a rather short lifetime of the proton 
compared with the supersymmetric
Standard Model, 
and the most parameter space can be explored by the future experiments such as the Hyper-Kamiokande 
experiment.
}

\end{center}
\end{titlepage}
\setcounter{footnote}{0}

\date{\today}

%%%%%%%%%%%%%%%%%%%%%%%%%%%%%%%%%%%%%%%%%%%%%%%%%%%%%%%%%%%%

%%%%%%%%%%%%%%%%%%%%%%%%%%%%%%%%%%%%%%%%%%%%%%%%%%%%%%%%%%%%
\section{Introduction}
After the discovery of the Higgs boson at the LHC experiments\,\cite{Aad:2012tfa}, 
the Standard Model (SM) was fully established as the unified theory of the weak and electromagnetic
interactions.
With the success of the unified electroweak theory, it is worthy to reappraise the long-sought 
idea, the grand unified theory (GUT)\,\cite{Georgi:1974sy} (for a review see e.g. \cite{Langacker:1980js}).
The GUT has  served as a guiding principle for theories beyond the Standard Model
where the three gauge coupling constants in the SM are unified into a universal gauge coupling constant
at a very high energy scale, i.e. the unification scale\,\cite{Georgi:1974yf}.
As is well known, however, the naive extrapolations of the three gauge coupling constants towards 
the high energy scale do not meet very precisely at a single scale~\cite{Amaldi:1991cn,Ellis:1990wk,Langacker:1991an}. 
Therefore, the ideas of the GUT inevitably require new charged particles with masses above the 
electroweak scale but below the unification scale.

Along with the GUT, dark matter has been another no less important guiding principle for theories beyond the SM. 
In fact, the cold dark matter paradigm is a pillar of  modern cosmology, whose 
existence has been established by numerous cosmological and astrophysical 
observations on a wide range of scales.
Although its detailed nature has remained unknown, we are almost certain 
that dark matter is not a part of the SM, and hence, its identification 
is the most important challenge in cosmology, astrophysics, and particle 
physics (for reviews, see e.g. \cite{Bertone:2004pz,Jungman:1995df,Murayama:2007ek}).

For decades, these two  principles have served as important criteria for assessing
how attractive a model of beyond the SM is.
For example, these two guiding principles are beautifully satisfied in the minimal supersymmetric 
Standard Model (MSSM).
With these successes, the MSSM remains one of the leading candidates of the theory beyond the SM,
despite the fact that the relatively large Higgs boson mass at around 125--126\,GeV and the null observations 
of the predicted superparticles seem to diminish one of the motivations of the MSSM, naturalness of the electroweak scale.%
\footnote{
The split supersymmetry~\cite{Giudice:2004tc,ArkaniHamed:2004yi} 
is a good example which weighs heavily on unification and dark matter rather than on
the naturalness.
See also e.g. Refs.\,\cite{Wells:2004di,Ibe:2006de,Acharya:2007rc,Hall:2011jd,Ibe:2011aa}
for less hierarchical but successful models with high scale supersymmetry breaking.
}

To achieve a model with a dark matter candidate and a precise unification
simultaneously, however, we do not need a large extension
of the SM as in the case of the MSSM, but it is possible with much smaller extensions.
For example, a smaller extension of the SM with only two Majorana Fermions in the adjoint
representations of $SU(2)_L$ and $SU(3)_c$ gauge groups of the SM is enough to 
achieved a precise coupling unification along with a good candidate for dark matter~\cite{Ibe:2009gt},
where their masses are required to be in the multi-TeV range and in the intermediate scale, respectively.%
\footnote{
For  generic discussion on the coupling unification by SM charged multiplets 
in the intermediate scale see \cite{Giudice:2012zp} (see also~\cite{Haba:2013via} for related discussion.)
}
Since the $SU(2)_L$ triplet Majorana fermion is predicted to be around the TeV scale,
the model is consistent with the so-called thermal ``minimal dark matter scenario"~\cite{Cirelli:2005uq} 
where the relic density of the thermally produced triplet fermion is consistent with the observation  
for its mass being $3$\,TeV~\cite{Hisano:2006nn,Cirelli:2005uq}.

In this paper, we revisit this small extension of the SM with the adjoint fermions 
as a low-energy effective theory below the GUT scale.
In particular, we discuss the non-thermal production of the triplet dark matter from the decay of the thermally
produced octet fermions whose mass is at around $10^{10-11}$\,GeV.
As we will show, the non-thermal contribution to the dark matter density dominates over the thermal contribution 
when the octet fermion decays through the higher dimensional operators suppressed by the GUT scale.
In such cases, the lighter triplet fermion than $3$\,TeV can also be a viable candidate for dark matter.
Due to the lighter mass of the triplet fermions, the model is more testable than the thermal minimal dark matter 
scenario in \cite{Cirelli:2005uq}. 
We also discuss how the dark matter mass is correlated to the proton lifetime predicted in the GUT models.

The organisation of the paper is as follows.
In section\,\ref{sec:unification}, we briefly review the small extension of the SM with 
adjoint Majorana fermions which allows a precise unification of the three gauge coupling 
constants at a high energy scale.
In section\,\ref{sec:DM}, we discuss the non-thermal production of the triplet fermion
from the decay of the thermally produced octet fermion.
There, we show that the non-thermally produced dark matter explains the observed dark matter density.
In section\,\ref{sec:proton decay}, we discuss the proton lifetime in this model.
The final section is devoted to our conclusions.

%%%%%%%%%%%%%%%%%%%%%%%%%%%%%%%%%%%%%%%%%
\section{Coupling Unification and Masses of Adjoint Fermions}
\label{sec:unification}
Let us briefly summarize the small extension of the SM with adjoint Majorana fermions.
In the following, we name the $SU(2)_L$ triplet and the $SU(3)_c$ octet Majorana fermions,
the wino-like fermion ($\tilde{w}$) and the gluino-like fermion ($\tilde{g}$), respectively, 
after the fashion of the MSSM.
Here, we discuss the masses of the adjoint fermions which allows a successful unification.
In this paper, we assume the minimal gauge group 
of the GUT to be $SU(5)$, where the leptons and quarks are unified into 
the $\bar{\bf 5}$ and $\bf 10$ representations respectively\,\cite{Langacker:1980js}.

To see how the three gauge coupling constants are extrapolated at high energy scales,
let us consider the renormalization group equations,
\begin{eqnarray}
 \frac{d\a_a^{-1}}{d \ln \m} = \frac{b_a}{2\pi}\ ,\,\,\, (a=1,2,3)\ ,
\end{eqnarray}
where $\m$ is the renormalization scale, and $\alpha_a = g_a^2/4\pi$ 
with $g_a$'s being the three gauge coupling constants of the SM.
The parameters $b_a$'s are so-called the coefficients of the beta functions. 
Since we are assuming the GUT with the $SU(5)$ gauge group,
we use the rescaled gauge coupling of the $U(1)_Y$ gauge interaction, {\it i.e.} $g_1 = \sqrt{5/3}\,g'$,

If the SM is an effective low energy theory of a GUT realized at a very high energy scale (the GUT scale), 
it is expected that the three gauge coupling constants meet together at around the GUT scale.
As is well known, however, the extrapolation of the SM gauge coupling constants
shows two failures of the SM as a low energy effective of a GUT;
\begin{itemize}
\item The coupling constants do not unify at a single scale, and hence, the unification is not precise at all.
\item If we take $g_1-g_2$ unification scale or $g_1-g_3$ unification scale as the GUT scale
which are at around $10^{13-14}$\,GeV,  
the predicted proton lifetime is too small to be consistent with the experimental constraints.
\end{itemize}
To circumvent these failures, we immediately find that there should be at least 
new fields charged under the $SU(3)_c$ and $SU(2)_L$ gauge groups to push up both the 
$g_1-g_{2}$ and $g_1-g_{3}$ unification scales while aiming at  precise unification.

%%%%%%%%%%%%%%%%%%%%%%%%%%%%%%%%%%%%%%%%%
\begin{figure}[t]
\begin{minipage}{.45\linewidth}
  \includegraphics[width=\linewidth]{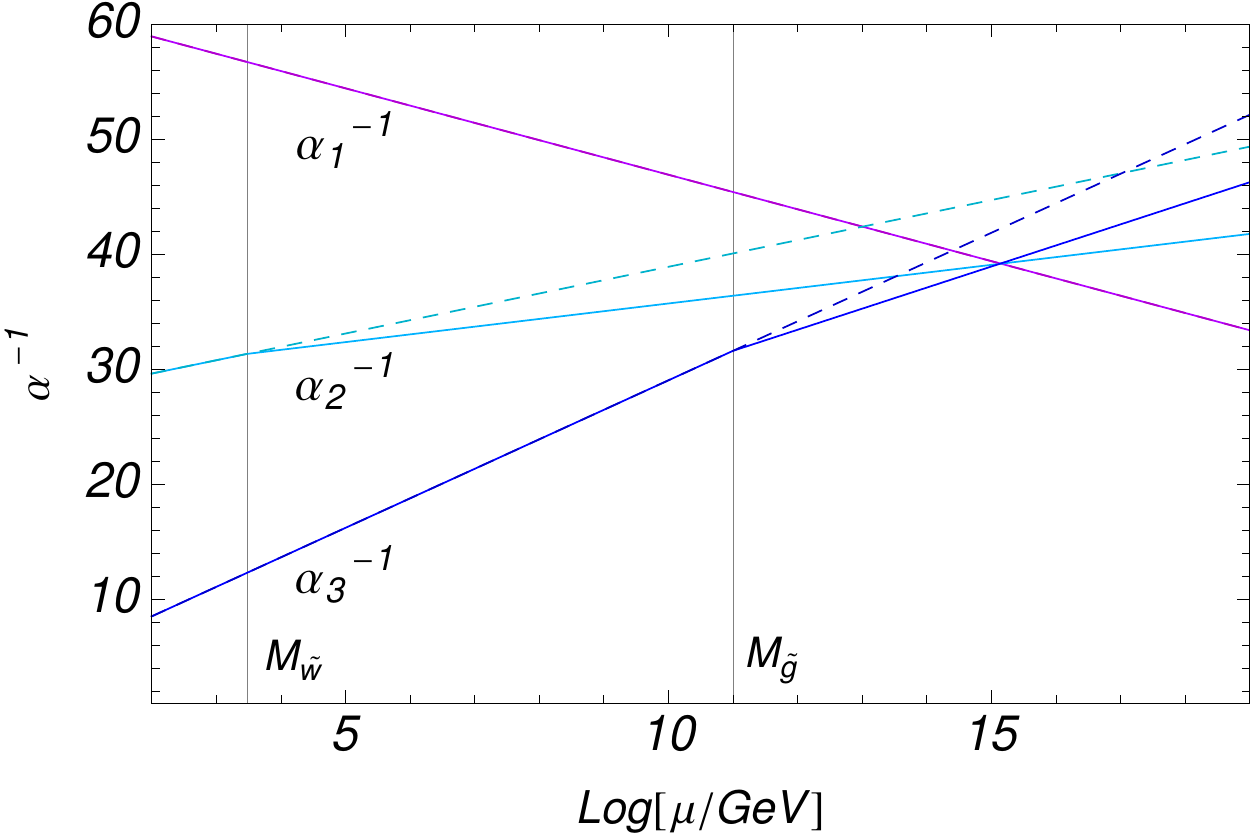}
 \end{minipage}
 \hspace{1cm}
 \begin{minipage}{.4\linewidth}
  \includegraphics[width=0.9\linewidth]{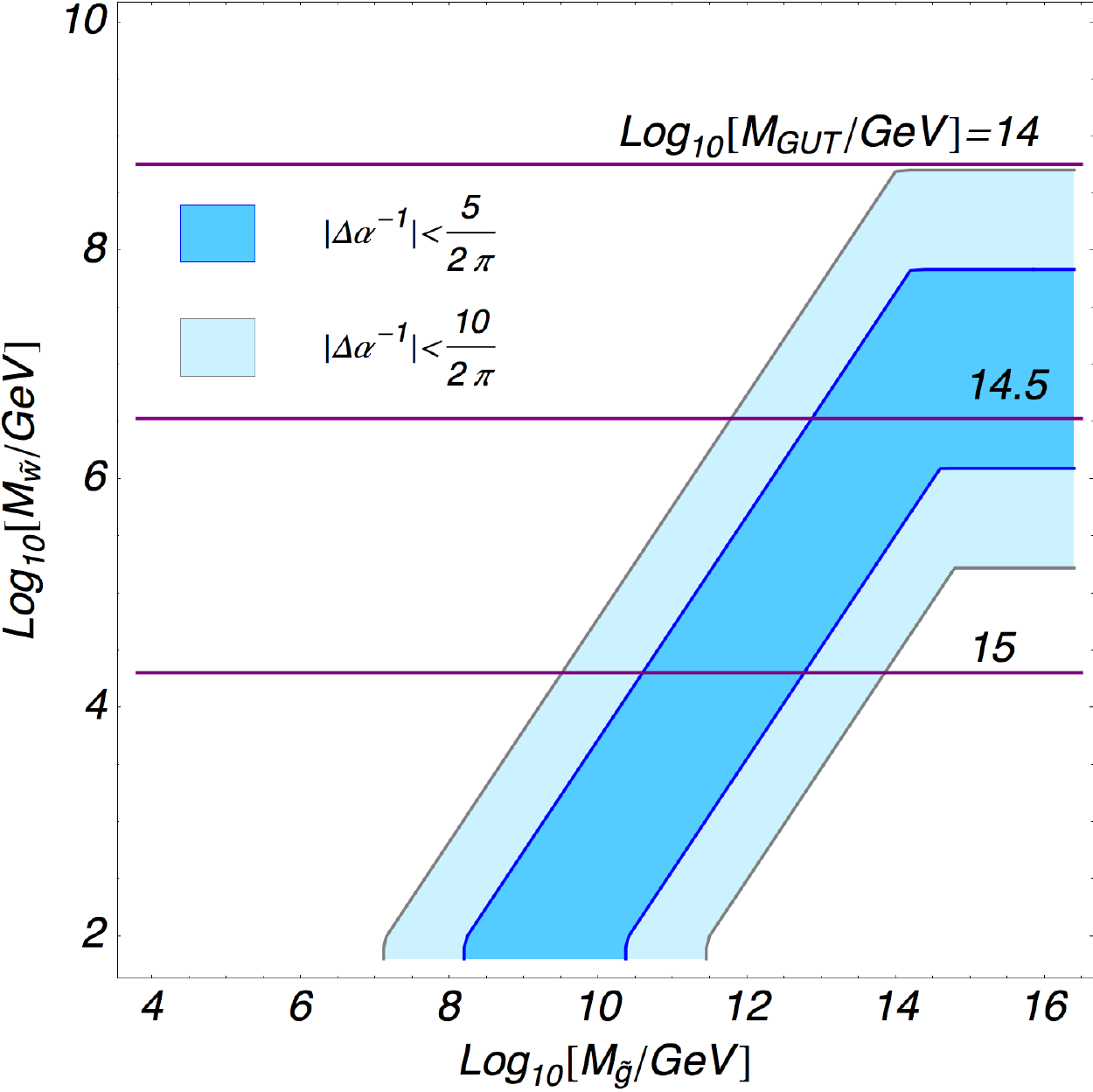}
 \end{minipage}
 \caption{\sl\small
(Left) An example of the one-loop renormalization group evolutions of the gauge coupling constants 
in the extended model (solid lines).
For a comparison, we show the evolutions in the SM as the dashed lines.
The masses of the wino-like and gluino-like fermions are taken to be $M_{\tilde w}=3$\,TeV 
and $M_{\tilde g}=10^{11}$\,GeV, respectively.
In the figures, we use $\alpha_3(m_Z)^{\overline{\rm MS}}=0.1185(6)$, $N_H =1$. 
We have also taken $m_{h}=126$\,GeV, and $m_{\rm top}=173.2$\,GeV,
although the results do not depend on these parameters significantly.
(Right) The adjoint fermion masses which are preferred for a precise unification.
In the (light-)blue shaded region, the three gauge coupling constants unify rather precisely, i.e. $|N_{\rm th}|<5(10)$,
respectively.
The horizontal lines show the contours of the  $g_1$--$g_2$ unification scale.
 }
\label{fig:unification}
\end{figure}
%%%%%%%%%%%%%%%%%%%%%%%%%%%%%%%%%%%%%%%%%

In Ref.\,\cite{Ibe:2009gt}, it was found that such a precise unification is achieved by introducing 
only two charged particles, one is an $SU(2)_L$ triplet Majorana fermion 
and the other is an $SU(3)_c$ octet Majorana fermion.
In this extended model, the coefficients  $b_a$'s at the one-loop level are given by 
\begin{eqnarray}
\label{eq:betas}
 b_1 =- \left(4+\frac{N_H}{10}\right)\ ,\quad
  b_2 = \left\{
\begin{array}{ccc}
\displaystyle{\frac{10}{3}-\frac{N_H}{6}}  &   (\m<M_{\tilde w})  \\
\displaystyle{ 2-\frac{N_H}{6}} &     (\m>M_{\tilde w})
\end{array}
\right.\ ,
\quad
 b_3 = \left\{
\begin{array}{ccc}
\displaystyle{7}  &   (\m<M_{\tilde g})  \\
\displaystyle{5} &     (\m>M_{\tilde g})
\end{array}
\right.\ ,
\end{eqnarray}
above the electroweak scale.
Here, $M_{\tilde w, \tilde g}$ denote the 
Majorana masses of the adjoint fermions.
In this paper, we fix the number of the Higgs doublet $N_H$ to be $N_H = 1$.
In the left panel of Fig.~\ref{fig:unification}, we show an example of the renormalization group evolutions 
of $\a_a^{-1}$ in the extended model at the one-loop level for $M_{\tilde w}=3$\,TeV and $M_{\tilde g}=10^{11}$\,GeV.
The figure shows that the three gauge coupling constants unify rather precisely at around $10^{15}$\,GeV
for these adjoint fermion masses.

Now, let us discuss the mass range of the adjoint fermions which is preferred for a precise unification.
For that purpose, we first need to quantify how precise the unification of the gauge coupling constants should be.
One caution here is that, for a given model of the GUT, the three gauge coupling constants 
in the effective low energy theory are defined by matching them to the universal gauge coupling constants
of the GUT by taking the threshold corrections from the charged particles with masses at the GUT scale
into account.
Therefore, the predictions on the adjoint fermion masses preferred by a precise unification 
inevitably depend on the details of the GUT models.

In this study, however, instead of specifying the GUT models, we quantify the degree of unification 
in terms of the size of the required threshold correction at the unification scale, 
so that the prediction becomes GUT model independent. 
Concretely, we define the unification scale $M_{\rm GUT}$ as the $g_1-g_2$
unification scale, and allows the adjoint fermion mass if the deviation of 
$g_3$ from $g_{1,2}$ at $M_{\rm GUT}$ is within some acceptance, 
\begin{eqnarray}
\label{eq:NTH}
N_{\rm th} \equiv 2\pi|{\mit\D} \alpha^{-1}| = 2\pi\times|\alpha_{\rm 1,2}^{-1}(M_{\rm GUT})- \alpha_{3}^{-1}(M_{\rm GUT})|
 < N_{\rm th}^{\rm (MAX)}\ .
 \end{eqnarray}
Here, the quantity $N_{\rm th}$ measures how large threshold corrections are required 
so that the three gauge coupling constants in the low energy effective theory are obtained
from a universal gauge coupling constant in the GUT. 
Very roughly speaking, it counts the (signed) number of charged fields in the GUT models
(in the unit of the fundamental representation) which contribute to the threshold corrections
at the GUT scale. 
For example, in the case of the MSSM, the threshold parameter satisfies $|N_{\rm th}|\lsim 5$\,\cite{Bagger:1995bw}
when the superparticles in the MSSM are at around the TeV scale.%
\footnote{The parameter $N_{\rm th}$ is related to the threshold parameter 
$\c_g$ in Ref.\,\cite{Bagger:1995bw} by $\c_g = N_{\rm th}/4\pi\times\a_{\rm GUT} $.}
In the followings, we take $N_{\rm th}^{\rm (MAX)} = 5(10)$ as the maximum acceptance 
so that the nominal unification in the adjoint extended model can be meaningfully interpreted 
as in the case of the MSSM.

In the left panel of Fig.\,\ref{fig:unification}, 
we show the degree of unification in the $M_{\tilde g}$--$M_{\tilde w}$ plane.
The figures show that a precise unification, $N_{\rm th}\lesssim 5$, is achieved for 
$M_{\tilde g} \simeq 10^{7} \times M_{\tilde w}$.
As we will discuss in section\,\ref{sec:proton decay}, the proton lifetime is predicted to be too short 
to be consistent with the current lower limit for $M_{\rm GUT} \lesssim 10^{15}$\,GeV.
Thus, by taking account of the proton lifetime, we find that a successful unification is achieved for
\begin{eqnarray}
\label{eq:unifMw}
  M_{\tilde w}  \simeq 10^{2-4}\,{\rm GeV}\ ,\quad {\rm and} \quad M_{\tilde g} \simeq 10^{6-8} \times M_{\tilde w}\ ,
\end{eqnarray}
in the adjoint extension model.

%%%%%%%%%%%%%%%%%%%%%%%%%%%%%%%%%%%%%%%%%%
\section{Non-thermal Minimal Dark Matter}
\label{sec:DM}
% Reviewing the minimal dark matter.
As we have seen in the previous section, a precise coupling unification is successfully achieved 
in the adjoint extension model for $M_{\tilde w} \simeq 10^{2-4}$\,GeV.
Interestingly, the wino-like fermion, i.e. the $SU(2)_L$ triplet Majorana fermion, has been considered 
as a good candidate for dark matter as the minimal dark matter model~\cite{Cirelli:2005uq}.%
\footnote{
The wino-like dark matter is also predicted in the MSSM with the anomaly mediated gaugino 
masses~\cite{Giudice:1998xp,Randall:1998uk,Dine:1992yw}.
(See Refs.\,\cite{Bagger:1999rd,D'Eramo:2013mya,Harigaya:2014sfa}
for the anomaly mediated gaugino masses in superspace formalism of supergravity.)
} 
With the rather large annihilation cross section, 
the observed dark matter density, $\Omega h^2 \simeq 0.1199\pm 0.0027$\,\cite{Ade:2013zuv},
can be achieved by its thermal relic density for $M_{\tilde w} \simeq 3$\,TeV~\cite{Hisano:2006nn,Cirelli:2005uq}.%
\footnote{See also \cite{Bauer:2014ula,Ovanesyan:2014fwa} 
for recent developments of the effective field theory approach to calculate
the relic density of the wino-like dark matter.}
The predicted relic density decreases quickly for a lighter wino-like fermion.

In this section, we discuss the non-thermal production of the wino-like dark matter.
The non-thermal contributions to the wino-like dark matter density have been
considered in the MSSM from the late-time decays of the moduli, the gravitino, 
and other 
sectors\,\cite{Moroi:1999zb,Gherghetta:1999sw,Ibe:2004tg,Ibe:2006de,Moroi:2013sla,Baer:2014eja,Blinov:2014nla}.
When the long-lived particles decay after the wino-like fermion has freezed-out 
from the thermal bath, the non-thermal contributions add up to the wino-like dark matter density.
With the non-thermal contributions, it is possible to explain the observed dark matter density 
with a wino-like fermion lighter than $3$\,TeV.

Interestingly, in the adjoint extended model, we already have a candidate for the source of the non-thermal contribution,
the gluino-like fermion.
The gluino-like fermion is expected to be in the thermal bath if the reheating temperature 
after inflation is much higher than the mass of the gluino-like fermion.
Such a high reheating temperature is preferred in thermal leptogenesis scenario~\cite{leptogenesis}.
As we will see shortly, the gluino-like fermion has an appropriate lifetime as a source 
of the non-thermal contribution when it decays through the higher dimensional operator suppressed
by the GUT scale.
Thus, in the adjoint extension model, the wino-like fermion with a mass smaller than $3$\,TeV is 
also a viable candidate for dark matter.

\subsection{Decay rate and abundance of the gluino-like fermion}
\label{sec:gluinodecay}
In our discussion, 
we have implicitly assumed that there is a symmetry which makes the wino-like fermion
stable so that the wino-like fermion can be a dark matter candidate.
Here, we take $Z_2$ symmetry as an example and assume 
that the adjoint fermions are odd under the $Z_2$-symmetry while other
SM particles are not charged under the symmetry.
In this case, the decay of the gluino-like fermion proceeds only through higher dimensional operators such as
\begin{eqnarray}
\label{eq:decay operator}
{\cal L} = \frac{1}{M_*^2} ({\tilde g}^A {t^A} {Q_L})^\dagger ({\tilde w}^I {\tau^I} {Q_L}) + h.c.\ ,
\end{eqnarray}
with a suppression scale $M_*$.
Here, $Q_L$ denotes a doublet quark in the SM. 
The Gell-Mann matrices $t^A$ and the Pauli matrices $\tau^I$ are normalized 
so that $\tr[t^At^B] = \delta^{AB}/2$ and  $\tr[\tau^I\tau^J] = \delta^{IJ}/2$.
Those decay operators can be, for example, generated  by exchanges of  
``squark-like fields" with masses of the order of the GUT scale or above, i.e, $M_{*} \gtrsim M_{\rm GUT}$.%
\footnote{For explicit calculation of the gluino decay via the squark exchange, see e.g.~\cite{Ibe:2012hr}.}
It should be noted that this assumption is quite consistent with 
the adjoint extension of the SM as a low energy effective theory of the GUT 
in which we do not require other fields than the ones in the SM and the adjoint fermions.%
\footnote{The precise unification in the previous section achieved by the adjoint fermion is still intact
even if there are ``squark-like fields" with masses below the GUT scale as long 
as they are also accompanied by "slepton-like fields" so that they form a complete 
multiplet of $SU(5)$, although we do not pursue such possibilities in this paper. 
}

%%%%%%%%%%%%%%%%%%%%%%%%%%%%%%%%%%%%%%%%%
\begin{figure}[t]
\begin{minipage}{.45\linewidth}
  \includegraphics[width=\linewidth]{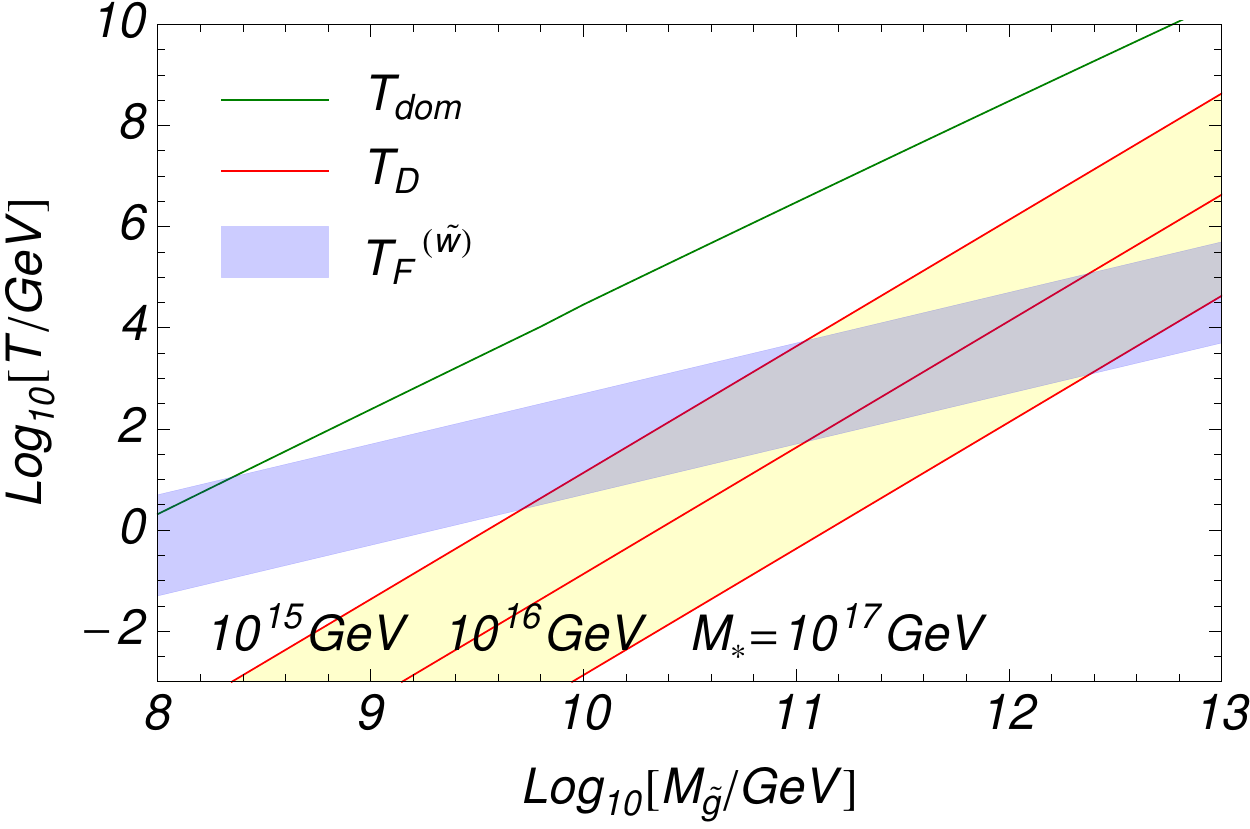}
 \end{minipage}
 \hspace{1cm}
 \begin{minipage}{.45\linewidth}
  \includegraphics[width=\linewidth]{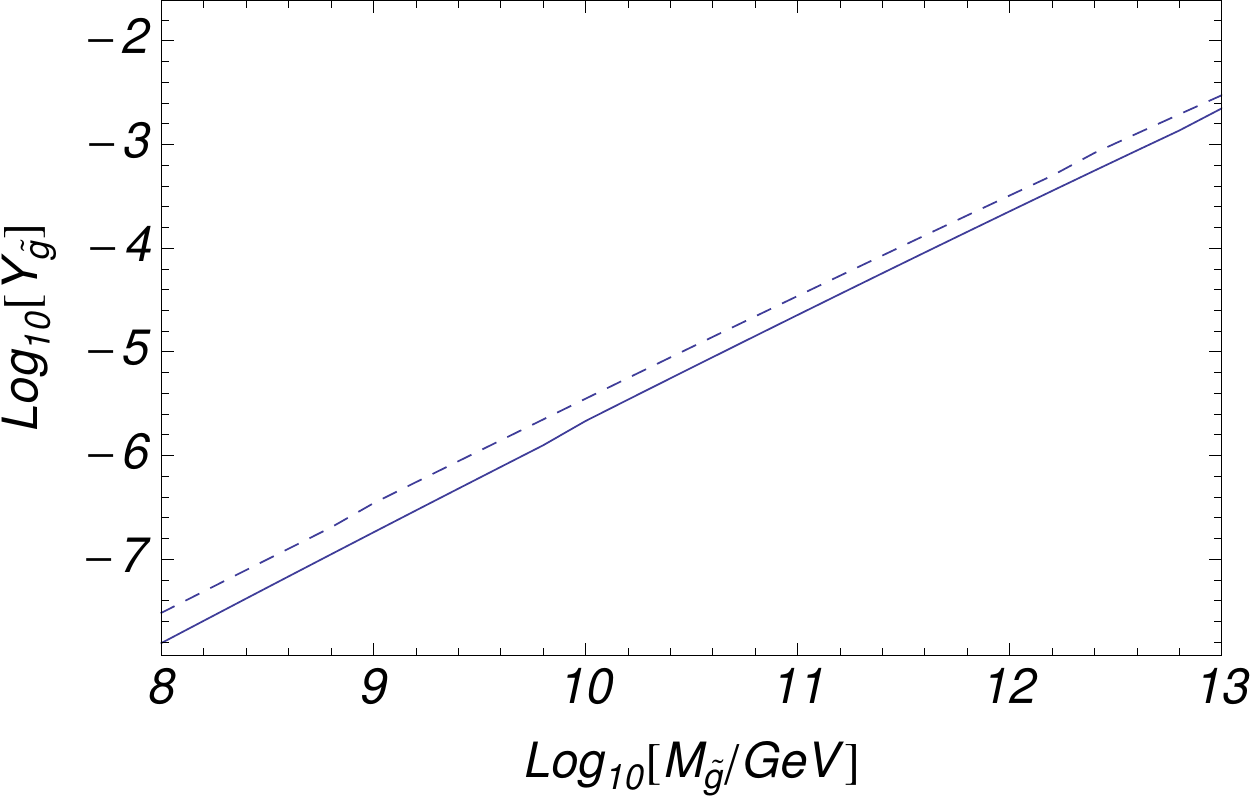}
 \end{minipage}
 \caption{\sl\small
(Left) The decay temperature of the gluino-like fermion for a given $M_*$ (Eq.\,(\ref{eq:TD})) (red lines)
and the dominate temperature (Eq.\,(\ref{eq:Tdom})) (green line).
For comparison, we also show a typical freeze-out temperature of the wino-like fermion,
$T_{F}(\tilde w)\simeq M_{\tilde w}/20$, as a blue shaded region assuming 
$M_{\tilde w} \simeq 10^{-(6-8)}M_{\tilde g}$ (Eq.\,(\ref{eq:unifMw})). 
(Right) The yield of the gluino after freeze-out (solid line).  For comparison,
 we show the yield without including the Sommerfeld enhancement factor as a dashed line. }
\label{fig:gluino}
\end{figure}
%%%%%%%%%%%%%%%%%%%%%%%%%%%%%%%%%%%%%%%%%

Through the operator in Eq.\,(\ref{eq:decay operator}), 
the gluino-like fermion decay into a pair of the doublet quarks and the wino-like fermion with a decay rate
roughly given by
\begin{eqnarray}
\Gamma_{\tilde g}  \simeq \frac{1}{(16\pi)^3} \frac{M_{\tilde {g}}^5}{M_*^4}\ .
\end{eqnarray}
The corresponding decay temperature defined by
\begin{eqnarray}
T_{D} \simeq \left(\frac{10}{\pi^2 g_*} M_{\rm PL}^2 \Gamma_{\tilde g}^2\right)^{1/4}\ ,
\end{eqnarray}
is estimated to be,
\begin{eqnarray}
\label{eq:TD}
T_{D} \simeq 44\,{\rm GeV}\times \left(\frac{111}{g_*}\right)^{1/4} \left(\frac{M_{\tilde g}}{10^{11}\,{\rm GeV}}\right)^{5/2}
\left(\frac{10^{16}\,\rm GeV}{M_{*}}\right)^{2}\ .
\end{eqnarray}
Here, $M_{\rm PL} \simeq 2.44 \times 10^{18}$\,GeV is the reduced Planck scale and 
we have fixed the effective number of the massless degrees of freedom to be $g_* \simeq 111$
which includes the SM particles and the wino-like fermions.

In the left panel of Fig.\,\ref{fig:gluino}, we show the decay temperature of the gluino-like fermion for 
given values of $M_*$.
For comparison, we also show a typical freeze-out temperature of the wino-like fermion,
$T_{F}^{(\tilde w)}\simeq M_{\tilde w}/25$, as a blue shaded region assuming 
$M_{\tilde w} \simeq 10^{-(6-8)}M_{\tilde g}$.
As a result, the decay temperature is expected to be below the 
freeze-out temperature for $M_* \gtrsim M_{\rm GUT}\simeq 10^{15}$\,GeV,
and hence, the gluino-like fermion in the intermediate scale can be a non-thermal source of the 
wino-like fermion.

Let us estimate the number density of the gluino-like fermion before its decay.
When the reheating temperature of the universe after inflation is much 
higher than the mass of the gluino-like fermion,  the gluino-like fermion 
is in the thermal bath.
The gluino-like fermion eventually decouples from the thermal bath when the cosmic temperature
decreases below the freeze-out temperature, $T_F^{(\tilde g)}$ which is given iteratively by
\begin{eqnarray}
\label{eq:xF}
\ln \left[ \frac{\vev{\s_{\tilde g} v}}{\pi^3}
\sqrt{\frac{45\pi}{g_*(T_F)}} M_{\rm PL}M_{\tilde g} g_{\tilde g} x_F^{1/2}\right]
 = x_F\ ,
\end{eqnarray}
where  $x \equiv M_{\tilde g}/T$ and $g_{\tilde g} = 2\times 8$~\cite{Gondolo:1990dk}.
Here, $\vev{\s_{\tilde g} v}$ denotes the thermally averaged cross section of the gluino-like fermion. 
The resultant relic density of the gluino-like fermion per the entropy density $s$ is then given by,
\begin{eqnarray}
\label{eq:Ygluino}
Y_{\tilde g} \simeq \sqrt\frac{45}{\pi }
\left(\int_{x_F}^{\infty} \frac{h_*(T)}{\sqrt{g_*(T)}} \frac{M_{\rm PL} M_{\tilde g}}{x^2} 
\vev{\s_{\tilde g} v} dx\right)^{-1}
\simeq  K^{-1}\times \sqrt\frac{45}{ g_*(T_F)\pi}
\frac{x_F}{\alpha_3^2}
\frac{M_{\tilde g}}{M_{\rm PL}}\ .
\end{eqnarray}
Here, $h_*(T)$  is the entropy degree-of-freedom counting factor, which is very close to
$g_*(T)$.
In the right panel of Fig.\,\ref{fig:gluino}, we show the yield of the gluino-like fermion after freeze-out.
In our analysis, we have used the annihilation cross section given in Ref.\,\cite{Harigaya:2014dwa}
where the Sommerfeld enhancement factors are taken into account (see also Ref.\,\cite{Baer:1998pg}).%
\footnote{
In the non-relativistic limit without the Sommerfeld enhancement factor, the annihilation cross section
is given by $\sigma_{\tilde g} v \simeq 63/32 \times \alpha_3^2/M_{\tilde g}^2$.
}
The final expression of Eq.\,(\ref{eq:Ygluino}) fairly reproduces our numerical result in Fig.\,\ref{fig:gluino}
for a numerical factor $K \simeq 10$.

%%%%%%%%%%%%%%%%%%%%%%%%%%%%%%%%%%%%%%%%%
\begin{figure}[t]
\begin{center}
  \includegraphics[width=.5\linewidth]{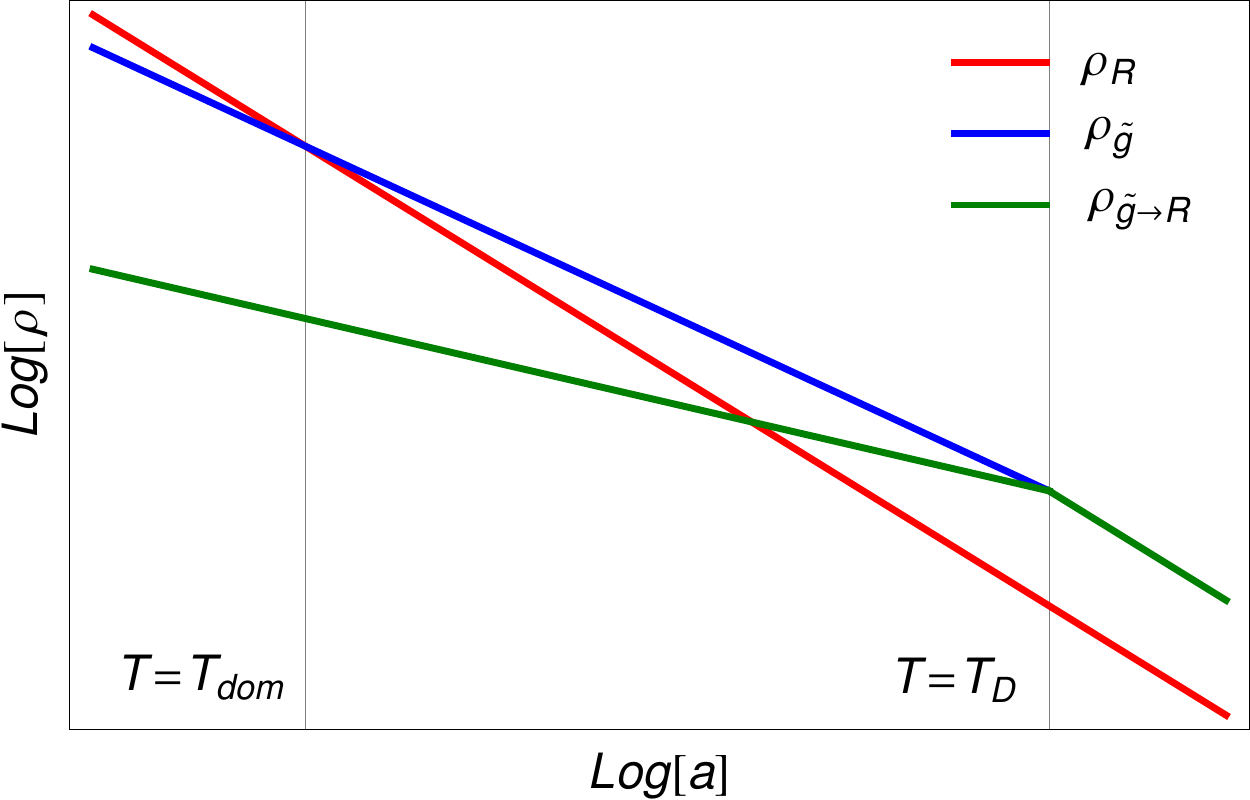}
 \caption{\sl\small
 The evolution of the energy densities of the radiation and the gluino-like fermion.
 Here, the energy density of the original radiation (red line) scales as $a^{-4}$ while the one 
 of the gluino-like fermion scales as $a^{-3}$ (blue line). 
 The radiation energy released by the occasional decay of the gluino-like fermion
 scales as $a^{-3/2}$ (green line).
 }
  \label{fig:en}
 \end{center}
\end{figure}
%%%%%%%%%%%%%%%%%%%%%%%%%%%%%%%%%%%%%%%%%

As the cosmic temperature further decreases, the energy density of the gluino-like fermion
becomes comparable to the radiation energy,
\begin{eqnarray}
\rho_{\rm tot}  = \rho_{R} + M_{\tilde g} Y_{\tilde g} s\ ,
\end{eqnarray}
and eventually dominates the energy density when the temperature becomes below $T_{\rm dom}$;
\begin{eqnarray}
\label{eq:Tdom}
T_{\rm dom} &\simeq& \frac{4}{3} M_{\tilde g} Y_{\tilde g}  
\simeq  \frac{4}{3K}\times \sqrt\frac{\pi}{45 g_*(T_F)}
\frac{x_F}{\alpha_3^2}
\frac{M_{\tilde g}^2}{M_{\rm PL}}\cr 
&\simeq& {3 \times 10^6}\,{\rm GeV}\left(
\frac{M_{\tilde g}}{10^{11}\,\rm GeV}
\right)^2.
\end{eqnarray}
By comparing with Eq.\,(\ref{eq:TD}), the domination temperature is higher than $T_{D}$ and $T_F^{({\tilde w})}$
in most parameter space, and hence, the gluino-like fermion decays after it dominates the energy density of the universe
(see the green-line in the left panel of Fig.\,\ref{fig:gluino}). 

In Fig.\,\ref{fig:en}, we show a schematic picture of the evolutions of the energy densities 
of the radiation and the gluino-like fermion after the freeze-out of the gluino-like fermion.
Once the domination by the gluino-like fermion happens, 
the energy density of the original radiation decreases and its temperature scales as $a^{-1}$
(the red line in the figure) with $a$ being the scale factor of the universe. 
At the same time, the gluino-like fermion gradually releases its energy into radiation  where 
the released radiation energy density scales as $a^{-3/2}$ (the green line in the figure).
Then, at around the decay temperature $T_D$, the most gluino-like fermion decays into 
radiation, after that its energy density decreases exponentially.
It should be noted that the scale factors at the domination time and the decay time are related to
the domination temperature and the decay temperature via,
\begin{eqnarray}
\left(\frac{a_{\rm dom}}{a_D}\right)^3 \simeq
\left(\frac{H_D}{H_{\rm dom}}\right)^2
\simeq
 \left(\frac{T_{D}}{T_{\rm dom} }\right)^4\ ,
\end{eqnarray}
where $H_{D}$ and $H_{\rm dom}$ denote the Hubble parameter at $T_{D}$ and $T_{\rm dom}$, respectively.
Thus, the radiation energy density is dominated by the one from the gluino-decay for
\begin{eqnarray}
\label{eq:Tstar}
T \lesssim T_* \equiv T_D \times \left(\frac{T_{\rm dom}}{T_D}\right)^{1/5} \ ,
\end{eqnarray}
although the final relic density of the wino-like fermion does not depend on the detailed evolution
of the thermal bath before $T_D$.

%%%%%%%%%%%%%%%%%%%%%%%%%%%%%%%%%%%%%%%%%%%%%%%%%%
\subsection{Non-thermal contribution to the relic density of the wino-like fermion}
Now, let us discuss how the wino-like fermion is produced  by 
the decay of the gluino-like fermion.
As we have discussed above, the gluino-like fermion decays through the higher dimensional 
operator which produces a wino-like fermion and a pair of quarks whose initial energies are of ${\cal O}(M_{\tilde g})$.
The high energetic quarks are immediately resolved into the thermal-bath and reach 
to the chemical equilibrium, which gradually heats up the radiation (see the appendix\,\ref{sec:thermalization}).
The produced charged components of the wino-like fermion also lose their energies 
immediately via the electromagnetic interactions with the 
thermal-bath~\cite{Ibe:2012hr}.
The neutral components of the wino-like fermion 
are on the other hand excited to the charged winos via inelastic scattering
and they lose energies as the charged components.%
\footnote{As we will see the required decay temperature to explain the observed dark matter density 
turns out to be $T_D > {\cal O}(1)$\,GeV, where the mass difference between
the neutral and the charged components around $160$\,MeV (see e.g.~\cite{Ibe:2012sx}) 
does not prevent the inelastic scattering.}

It should be noted that the wino-like fermions are also produced by high energy injection of 
the quarks into the thermal bath~\cite{Harigaya:2014waa}.
The production cross section of the wino-like fermion via  interactions between
the injected quarks and the quarks in the thermal-bath are roughly given by,
\begin{eqnarray}
\sigma_{\tilde w} \simeq \frac{\alpha_2 \alpha_3}{ET}\ .
\end{eqnarray}
Here, $E$ is the energy of the quarks/gluons which are induced 
during the thermalization process of the injected quarks from the decay of the gluino-like fermion.
The number of those energetic particles per a decay of the gluino-like fermion is 
given by,
\begin{eqnarray}
N(E)\sim \frac{M_{\tilde g}}{E}\ .
\end{eqnarray}
The energy loss rate of the quarks/gluons with energy $E$ via inelastic soft scattering by the 
QED interaction is, for example, given by~\cite{Harigaya:2013vwa} (see also \cite{Kurkela:2011ti}),
\begin{eqnarray}
\G_{\rm split}(E) \sim \alpha_3^2 T\sqrt{\frac{T}{E}}\ . 
\end{eqnarray}
Thus, the probability of the pair production of the wino-like fermion
from each particle with an energy $E$ is given by,
\begin{eqnarray}
P_{\tilde w} = {\min}\left[1,\sigma_{\tilde w}T^3  \G_{\rm split}^{-1}\right] \sim 
{\min}\left[1, \frac{\a_2}{\a_3}
\sqrt{\frac{T}{E}} \right] \ .
\label{eq:prob}
\end{eqnarray}
This probability is maximized for $E = E_{\rm th} \sim M_{\tilde w}^2/T$ for $T \ll M_{\rm w}$.
As a result, for $T \ll M_{\tilde w}$, the number of the wino-like fermion produced by 
a decay of the gluino-like fermion is given by,
\begin{eqnarray}
N_{\tilde w} = {\max}\left[1, N( E_{\rm th}) \times P_{\tilde w}\right]\ .
\label{eq:Ntildew}
\end{eqnarray}
In our model, we find that $N_{\tilde w}\gg1$ 
in most parameter space for $T_D \gtrsim 1$\,GeV, and hence, 
the wino-like fermion produced by the gluino-decay is dominated by the contribution
from the secondary generation.
As we will see, the final relic abundance of the wino-like fermion, however, does not depend on $N_{\tilde w}$
significantly as long as $N_{\tilde w} \gtrsim 1$
and the final abundance is mainly determined by the annihilation rate
of the wino-like fermion at the time of $T\simeq T_D$.

%%%%%%%%%%%%%%%%%%%%%%%%%%%%%%%%%%%%%%%%%%%%%%%%%
\subsection{Relic density of the wino-like fermion}
Let us estimate the relic density of the wino-like fermion by solving the set of the Boltzmann equations, 
\begin{eqnarray}
\frac{dn_{\tilde{w}}}{dt} + 3 H n_{\tilde{w}} &=& - \vev{\s_{\mathrm{eff}}v}(n^2_{\tilde{w}}-n^2_{\tilde{w},\mathrm{eq}}) + N_{\tilde{w}} \Gamma_{\tilde{g}} n_{\tilde{g}}\ , 
\label{eq:Bwino}\\
\frac{dn_{\tilde{g}}}{dt} + 3 H n_{\tilde{g}} &=& -\Gamma_{\tilde{g}} n_{\tilde{g}}\ , \label{eq:Bgluino}\\
\frac{d\r_\mathrm{R}}{dt} + 4 H \r_\mathrm{R} &=& (M_{\tilde{g}}-N_{\tilde{w}} M_{\tilde{w}}) \G_{\tilde{g}} n_{\tilde{g}} + M_{\tilde{w}} \vev{\s_{\mathrm{eff}}v} n^2_{\tilde{w}} \ .
\label{eq:injection}
\end{eqnarray}
Here, $n_{\tilde{w}}$ and $n_{\tilde{g}}$ are the number densities of wino-like fermions and gluino-like fermions, 
respectively, $n_{\tilde{w},\rm eq}$ the thermal-equilibrium value of $n_{\tilde{w}}$~\cite{Moroi:2013sla}.
In Eq.\,(\ref{eq:injection}), the heat injection from the wino-like fermions is proportional to 
$M_{\tilde w}$, since the wino-like fermions loose their energy into the thermal bath immediately after the production. 
Accordingly, the Hubble parameter $H$ during the non-thermal production period is approximated by,
\begin{eqnarray}
H^2 = \frac{1}{3M_{\rm PL}^2} \left(M_{\tilde g} n_{\tilde g} + \r_R  \right)\ ,
\end{eqnarray}
which is eventually dominated by the contributions from the radiation energy for $T \lesssim T_D$.
It should be noted that the thermally averaged effective annihilation cross section,
$\vev{\s_{\mathrm{eff}}v}$, is affected by the coannihilation process and the Sommerfeld enhancement,
which significantly enhances the cross section compared with the one predicted by tree-level contributions 
in the case of wino-like $SU(2)_L$ adjoint fermion~\cite{Hisano:2006nn}.
In our numerical calculation, we have taken into account those enhancement factors according to
\cite{Hisano:2006nn,Moroi:2013sla}.  
  
In order to obtain an approximated relic abundance, let us first consider the 
situation where all gluino-like fermions decay into wino-like fermion instantaneously at $T_{\mathrm{eff}}\simeq T_{D}$.
In this case, Eq.\,(\ref{eq:Bwino}) is reduced to
\begin{eqnarray}
\frac{dn_{\tilde{w}}}{dt} + 3 H n_{\tilde{w}} &=& - \vev{\s_{\mathrm{eff}}v}n^2_{\tilde{w}}\ , 
\label{eq:Bwino2}
\end{eqnarray}
for $T\ll T_{\rm eff}$ with $H$ dominated by the radiation contribution.
The solution of this reduced equation is given by,
\begin{eqnarray}
\left. \frac{1}{Y_{\tilde{w}}}\right|_{T\rightarrow 0} = \frac{1}{Y_{\tilde{w}}^{\rm(init)}} + \int_0^{T_{\mathrm{eff}}} \frac{\vev{\s_{\mathrm{eff}}v} s}{H T} dT\ ,
\label{eq:Bsol}
\end{eqnarray}
where $Y_{\tilde{w}}^{\rm(init)}$ accounts for the non-thermal production by the decay of the gluino-like fermions
at $T_{\rm eff}$, i.e.
\begin{eqnarray}
Y_{\tilde{w}}^{\rm(init)} \simeq \left. \frac{N_{\tilde w} n_{\tilde g}}{s}\right|_{T \simeq T_{\rm eff}}
\simeq N_{\tilde w}\frac{3T_{\rm eff}}{4 M_{\tilde g}} \ ,
\end{eqnarray}
where we have neglected thermally produced component and assumed that $T_{\rm eff} < T_{F}^{(\tilde w)}$.
By remembering that $N_{\tilde w} \gg 1$, we find that $Y_{\tilde{w}}^{\rm(init)}$ is much larger than 
the inverse of the second term of Eq.\,(\ref{eq:Bsol}), 
\begin{eqnarray}
Y_{\tilde{w}}^{\rm (asym)} \simeq \left(\int_0^{T_{\mathrm{eff}}} \frac{\vev{\s_{\mathrm{eff}}v} s}{H T} dT\right)^{-1}\ ,
\label{eq:Bsol2}
\end{eqnarray}
which can be further reduced 
\begin{eqnarray}
\bar{Y}_{\tilde{w}}^{\rm (asym)} \simeq \left. \frac{H}{\vev{\s_{\mathrm{eff}}v} s}\right|_{T\simeq T_{\mathrm{eff}}}\ .
\label{eq:Bsol3}
\end{eqnarray}
when $\vev{\s_{\mathrm{eff}}v} $ does not depend on temperature.

%%%%%%%%%%%%%%%%%%%%%%%%%%%%%%%%%%%%%%%%%
\begin{figure}[t]
\begin{minipage}{.45\linewidth}
  \includegraphics[width=\linewidth]{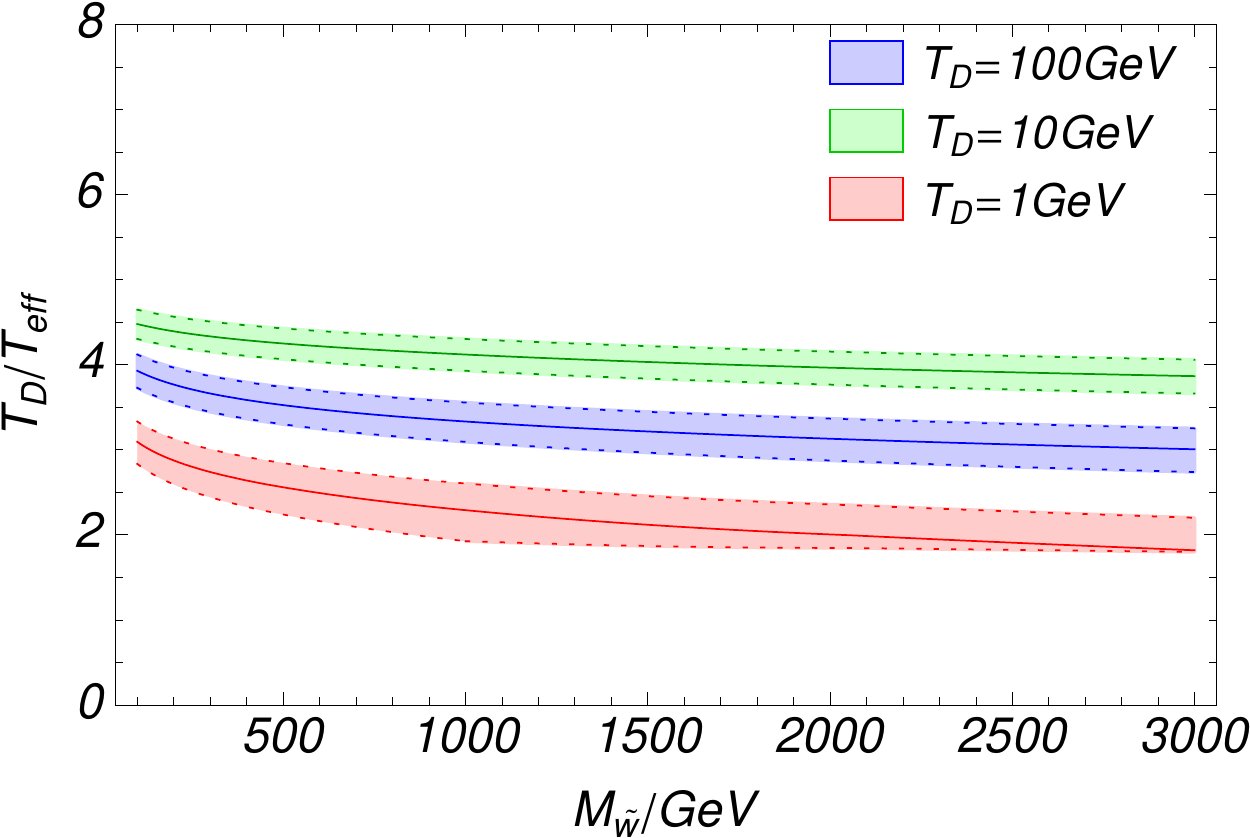}
 \end{minipage}
 \hspace{1cm}
 \begin{minipage}{.4\linewidth}
  \includegraphics[width=\linewidth]{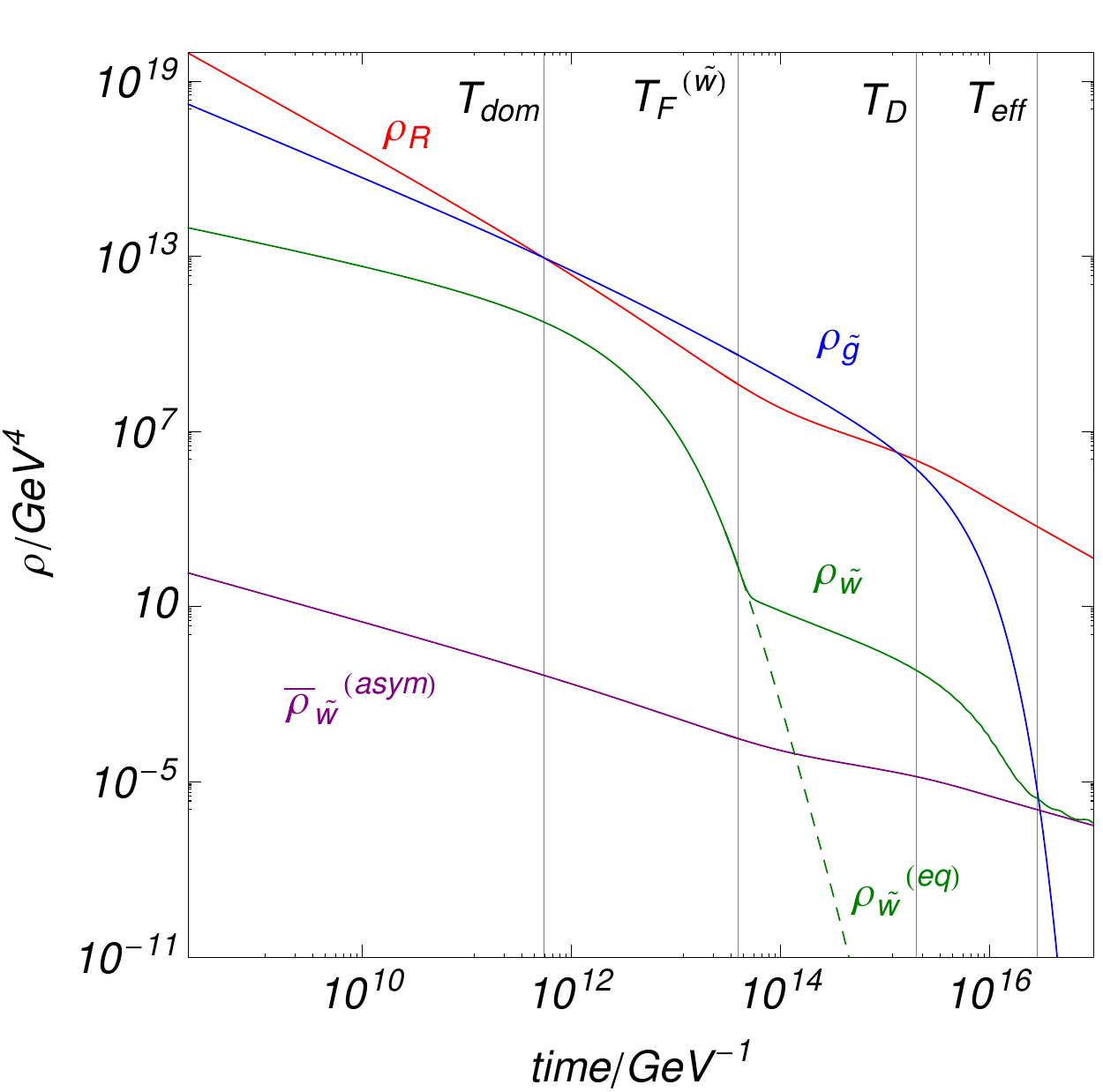}
 \end{minipage}
 \caption{\sl\small
(Left) 
The ratio between $T_D$ and the effective temperature of the non-thermal production $T_{\rm eff}$
for given $T_D$.
The solid lines show the ratios for $P_{\tilde w}$ given in Eq.\,(\ref{eq:prob}).
For comparisons, the ratios for ten times smaller (larger) values of $P_{\tilde w}$ than the 
one given in Eq.\,(\ref{eq:prob}) are shown as the lower (upper) dotted lines.
(Right) 
The evolutions of the energy densities for $M_{\tilde w} = 1$\,TeV, 
$M_{\tilde g} = 10^{10}$\,GeV, and $M_* = 10^{15}$\,GeV
obtained by solving the Boltzmann equations numerically.
 }
\label{fig:winoNumerical}
\end{figure}
%%%%%%%%%%%%%%%%%%%%%%%%%%%%%%%%%%%%%%%%%

In reality, the non-thermal production is not an immediate process.
However, by appropriately relating the effective temperature $T_{\rm eff}$
to the decay temperature $T_D$, we can obtain a good approximation 
of the yield of the wino-like fermion by using the above simplified solution in Eq.\,(\ref{eq:Bsol2}).
To find the relation between $T_D$ and $T_{\rm eff}$, 
let us look at the Boltzmann Eq.\,(\ref{eq:Bgluino}) at the temperature just below $T_D$; 
\begin{eqnarray}
\frac{dn_{\tilde{w}}}{dt} + 3 H n_{\tilde{w}} &=& - \vev{\s_{\mathrm{eff}}v}n^2_{\tilde{w}} + N_{\tilde{w}} \Gamma_{\tilde{g}} \frac{\r_{\tilde{g}}}{M_{\tilde{g}}} e^{- \G_{\tilde{g}} t }\ .
\end{eqnarray}
By assuming the radiation domination at that period, this equation can be rewritten in terms of the temperature,
\begin{eqnarray}
H T \frac{dY_{\tilde{w}}}{dT} = \vev{\s_{\mathrm{eff}}v} s Y^2_{\tilde{w}} - N_{\tilde{w}} \Gamma_{\tilde{g}} \frac{3T}{4 M_{\tilde{g}}} e^{- \frac{3}{2}\left(\frac{T_{D}}{T}\right)^2}.
\end{eqnarray}
Since $T_{\mathrm{eff}}$ can be regarded as the temperature at which the source term is dumped 
and becomes comparable to the first term, we find the relation between $T_{\rm eff}$ and $T_D$ by,
\begin{eqnarray}
T_{\mathrm{eff}} \simeq \sqrt{\frac{3}{2}}\, T_D \times 
\left[ \log
\left[{\frac{3N_{\tilde{w}} Y_0}{\bar{Y}_{\tilde{w}}^{\rm (asym)}}}\right]  \right]^{-\frac{1}{2}},
\label{eq:TDTeff}
\end{eqnarray}
where $Y_0 = 3T/4M_{\tilde g}$, and we have used $\G_{\tilde g} \simeq 3H$ at $T \simeq T_{D}$.

In the left panel of Fig.\,\ref{fig:winoNumerical}, we show the relation between $T_D$ and $T_{\rm eff}$ 
as a function of $M_{\tilde w}$ for given values of $T_D$.
In the figure, the solid lines show the ratios for $P_{\tilde w}$ given in Eq.\,(\ref{eq:prob}).
For comparisons, we also show the ratios for ten times smaller (larger) values of $P_{\tilde w}$ than the 
one given in Eq.\,(\ref{eq:prob}) as the dotted lines.
The figure shows that the ratios do not depend on the values of $P_{\tilde w}$ significantly.
It should be also noted that we have taken $M_{\tilde g} = 10^7\times M_{\tilde w}$ in the figure,
although $M_{\tilde g}$ dependence is not significant
since $N_{\tilde w}$ is proportional to $M_{\tilde w}$ while $Y_0$ is inversely proportional to $M_{\tilde g}$
in most parameter region.
In the right panel of  Fig.\,\ref{fig:winoNumerical}, we also show an example of the evolutions 
of the energy densities obtained by the Boltzmann equation numerically.
The figure shows that the energy density is dominated by gluino-like fermions at $T_{\rm dom}$.
After that the wino-like fermions decouple from the thermal bath at around $T^{(\tilde w)}_F \simeq m_{\tilde {w}}/25$.%
\footnote{The yield of the wino-like fermion at this point
is much higher than the one predicted in the case of the thermal freeze-out without the injection
from the decay of the gluino-like fermion.}
At the decay temperature $T_D$, the gluino-like fermion decays and the energy density gets dominated 
by the radiation energy.
Finally, the energy density of the wino-like fermion approaches to its asymptotic value,%
\footnote{
In Fig.\,\ref{fig:winoNumerical} , we instead show 
$\bar{\rho}^{\rm (asym)}_{\tilde w} = M_{\tilde w} H/\vev{\s_{\rm eff} v} $ in Fig.\,\ref{fig:winoNumerical}
as the purple line which coincides with $\rho^{\rm (asym)}_{\tilde w}$ for $T<T_{\rm eff}$.
}
\begin{eqnarray}
 \rho^{\rm (asym)}_{\tilde {w}} = M_{\tilde w}Y_{\tilde w}^{\rm (asym)} \times s(T) 
 \ .
\end{eqnarray}
The figure shows that the asymptotic solution in Eq.\,(\ref{eq:Bsol2}) gives a fairly good 
approximation. 

%%%%%%%%%%%%%%%%%%%%%%%%%%%%%%%%%%%%%%%%%
\begin{figure}[t]
\begin{minipage}{.45\linewidth}
  \includegraphics[width=\linewidth]{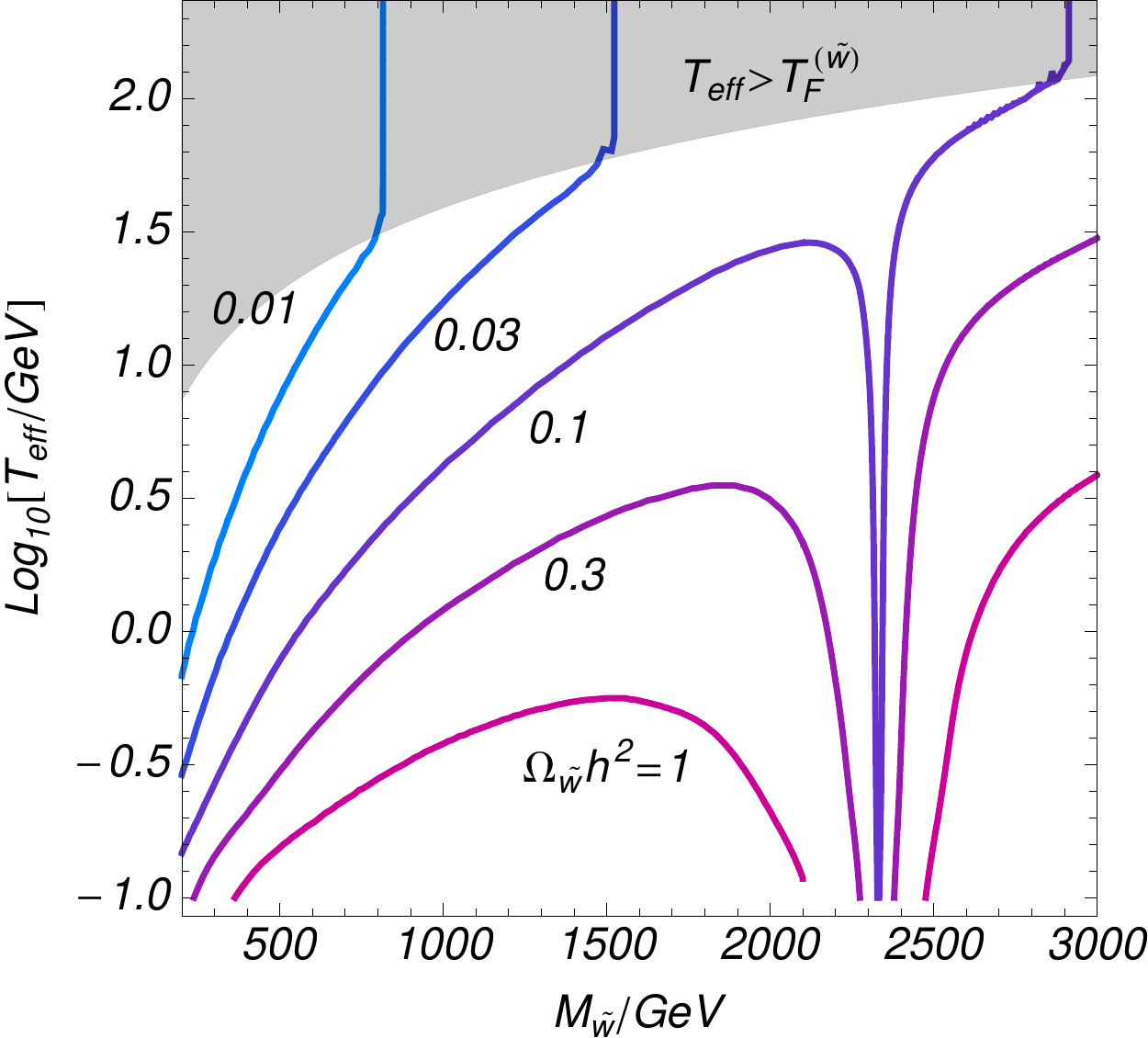}
 \end{minipage}
 \hspace{1cm}
 \begin{minipage}{.45\linewidth}
  \includegraphics[width=\linewidth]{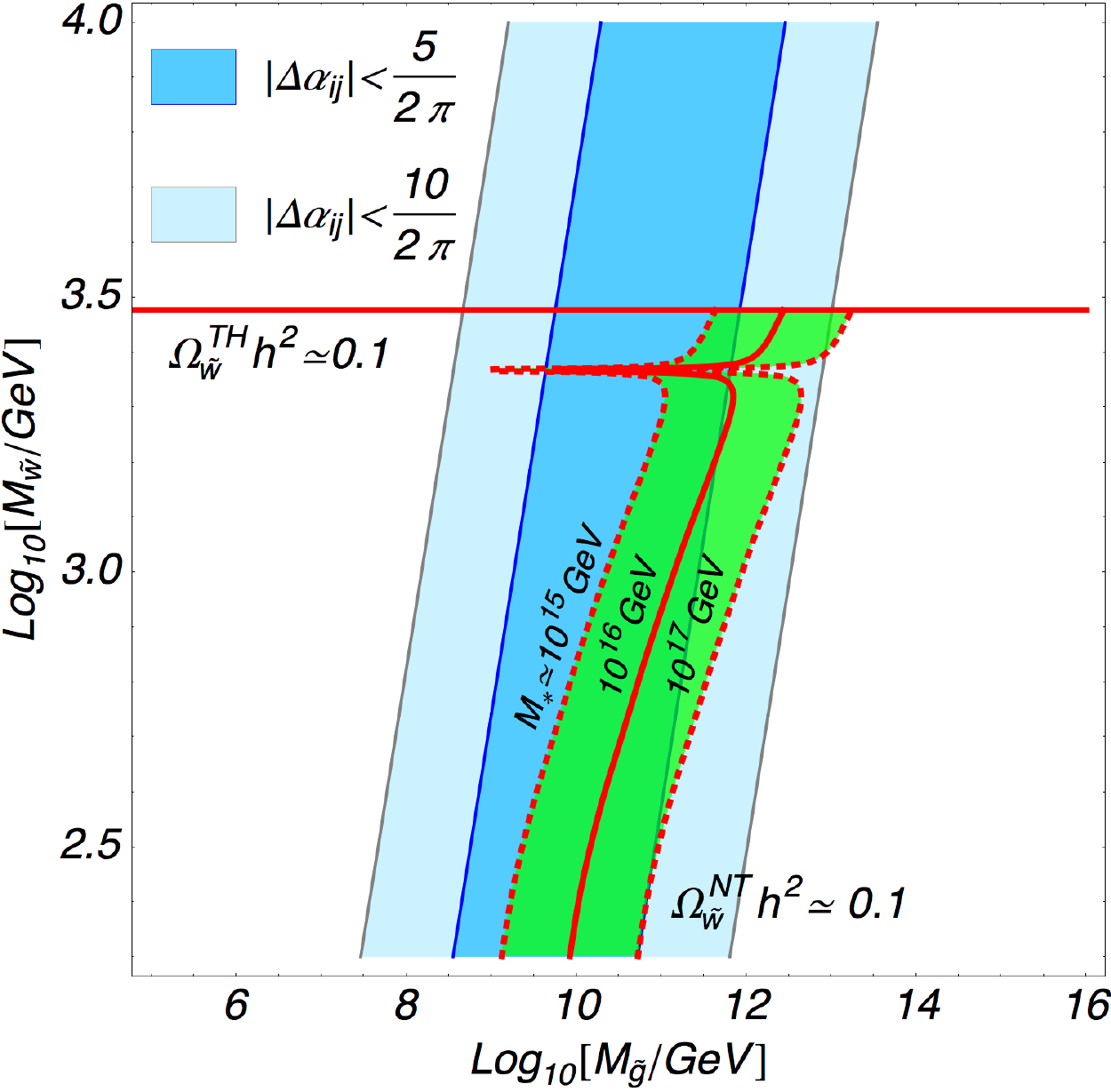}
 \end{minipage}
 \caption{\sl\small
(Left) Contour plots of $\Omega_{\tilde w}h^2$ on the $T_{\rm eff}$-$M_{\tilde w}$ plane.
(Right) Contour plots of $M_*$ which provides an appropriate decay temperature of the gluino-like fermion
for $\Omega_{\tilde{w}}h^2 \simeq 0.1146$ via the non-thermal production.
Here, 
we put the superscripts $\Omega_{\tilde{w}}^{TH}h^2 \simeq 0.1$ (thermal) and  
$\Omega_{\tilde{w}}^{NT}h^2 \simeq 0.1$ (non-thermal) to indicate the main contributions 
to the dark matter abundance.
The precise unification conditions in Fig.\,\ref{fig:unification} are also shown.
 }
\label{fig:wino}
\end{figure}
%%%%%%%%%%%%%%%%%%%%%%%%%%%%%%%%%%%%%%%%%

In the left panel of Fig.\,\ref{fig:wino}, we show the contours of 
$\Omega_{\tilde w} h^2$ in the $T_{\rm eff} - M_{\tilde w}$ plane. 
For a given mass of the wino-like fermion, the relic abundance becomes insensitive to $T_{\rm eff}$
when $T_{\rm eff}$ is so high that the wino-like fermion is still in thermal equilibrium after $T_{\mathrm{eff}}$. 
In such region, the relic abundance is dominated by thermal relic. 
The abundance of the wino-like fermion is significantly suppressed at 
around $M_{\tilde w} \sim 2300~ \mathrm{GeV}$ where the annihilation cross section
of the wino-like fermion is significantly enhanced by the Sommerfeld enhancement. 
We see from the left panel of Fig.\,\ref{fig:wino} that the region with a wino-like fermion mass smaller than $3$\,TeV 
can be consistent with the observed dark matter abundance for $T_{\rm eff} = {\cal O}(1$--$10)$\,GeV.%
\footnote{Due to $N_{\tilde w}\gg 1$, the required decay temperatures for a given value of $\Omega h^2$
are different from the ones given in \cite{Moroi:2013sla}.}

From Eqs.\,(\ref{eq:TD})
and (\ref{eq:TDTeff}), we can estimate the required size of $M_*$ to achieve an appropriate decay temperature
for $\Omega_{\tilde w}h^2 \simeq 0.1$
as a function of $M_{\tilde w}$ and $M_{\tilde g}$.
In the left panel of Fig.\,\ref{fig:wino}, we  show the contour plot of the required $M_*$ 
by overlaying the left panel of Fig.\,\ref{fig:unification}.
The figure shows that in the mass region which leads to the appropriate decay temperature 
for $M_* \simeq 10^{15-17}$\,GeV is consistent with the successful unification.
Therefore, we find that gluino-like fermion can be a successful source of the non-thermal
wino-like fermion without requiring lighter fields than the GUT scale.

\section{Constraints from Proton Decay}
\label{sec:proton decay}
In section \ref{sec:unification}, we discussed the mass range of the adjoint fermions 
which leads to a precise unification of the three gauge coupling constants.
There, we  found that the GUT scale (defined as a $g_1-g_2$ unification scale)
has a strong correlation with the mass of the wino-like fermion.
The GUT scale is, in turn, correlated with the mass of the massive gauge bosons
in the GUT (i.e. the GUT gauge boson),  which causes a decay of the proton~\cite{Georgi:1974sy,Georgi:1974yf}.
In this section, we discuss the proton lifetime expected in the adjoint extension model.

To estimate the mass of the GUT gauge bosons, let us review the matching conditions of the 
three gauge coupling constants to the universal gauge coupling constant of the GUT at the one-loop level;
\begin{eqnarray}
\frac{1}{\alpha_3(\mu)} &=& \frac{1}{\alpha_{\rm GUT}(\Lambda)} + \frac{1}{2\pi} 
\left(
\beta_{3}^{(L)} \log\frac{\mu}{\Lambda}
+
\beta_{3}^{(XY)} \log\frac{M_{V}}{\Lambda}
+
{\mit \D}\beta_{3} \log\frac{{\cal M}_{3}}{\Lambda}
\right)\ , \nonumber\\
\frac{1}{\alpha_2(\mu)} &=& \frac{1}{\alpha_{\rm GUT}(\Lambda)} + \frac{1}{2\pi} 
\left(
\beta_{2}^{(L)} \log\frac{\mu}{\Lambda}
+
\beta_{2}^{(XY)} \log\frac{M_{V}}{\Lambda}
+
{\mit \D}\beta_{2} \log\frac{{\cal M}_{2}}{\Lambda}
\right)\ , \nonumber\\
\frac{1}{\alpha_1(\mu)} &=& \frac{1}{\alpha_{\rm GUT}(\Lambda)} + \frac{1}{2\pi} 
\left(
\beta_{1}^{(L)} \log\frac{\mu}{\Lambda}
+
\beta_{1}^{(XY)} \log\frac{M_{V}}{\Lambda}
+
{\mit \D}\beta_{1} \log\frac{{\cal M}_{1}}{\Lambda}
\right)\ .
\end{eqnarray}
Here, $M_V$ denotes the mass of the GUT gauge boson, $\mu$ the renormalisation scale,
and $\Lambda$ the scale at which the boundary condition of $\alpha_{\rm GUT}$ is given.
The  coefficients of the beta function are given by
\begin{eqnarray}
\label{eq:beta_GUT}
\beta_{3}^{(L)} &=& 5\ , \quad
\beta_{2}^{(L)} = \frac{11}{6} \ , \quad
\beta_{1}^{(L)} = - \frac{41}{10}\ ,  \nonumber\\
\beta_{3}^{(XY)} &=& 7\ , \quad
\beta_{2}^{(XY)} = \frac{21}{2} \ , \quad
\beta_{1}^{(XY)} = - \frac{25}{2}\ , 
\end{eqnarray}
respectively.
The final terms in each matching condition collectively denote the contributions from other multiplets 
at the GUT scale than the ones in the SM, the adjoint fermions, and the GUT gauge bosons, 
which depend on the details of the GUT models.

\begin{figure}[t]
\begin{minipage}{.45\linewidth}
  \includegraphics[width=\linewidth]{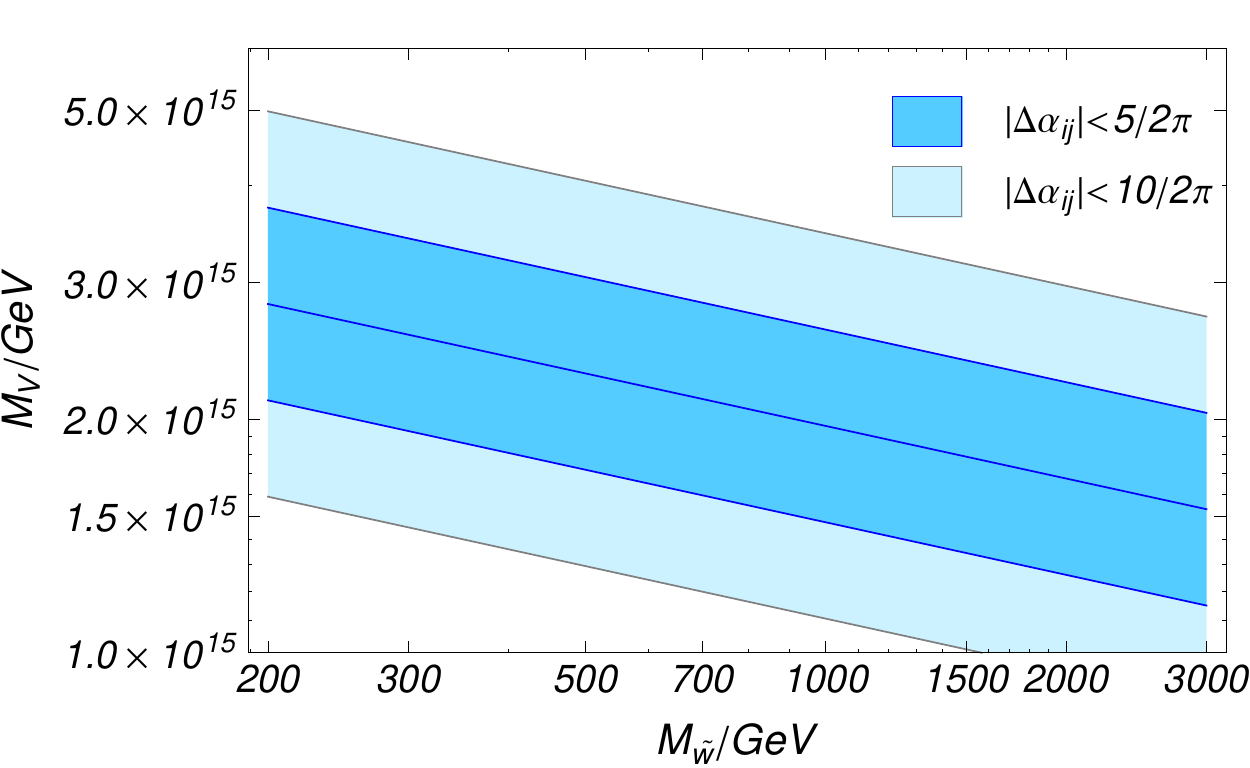}
 \end{minipage}
 \hspace{1cm}
 \begin{minipage}{.45\linewidth}
  \includegraphics[width=\linewidth]{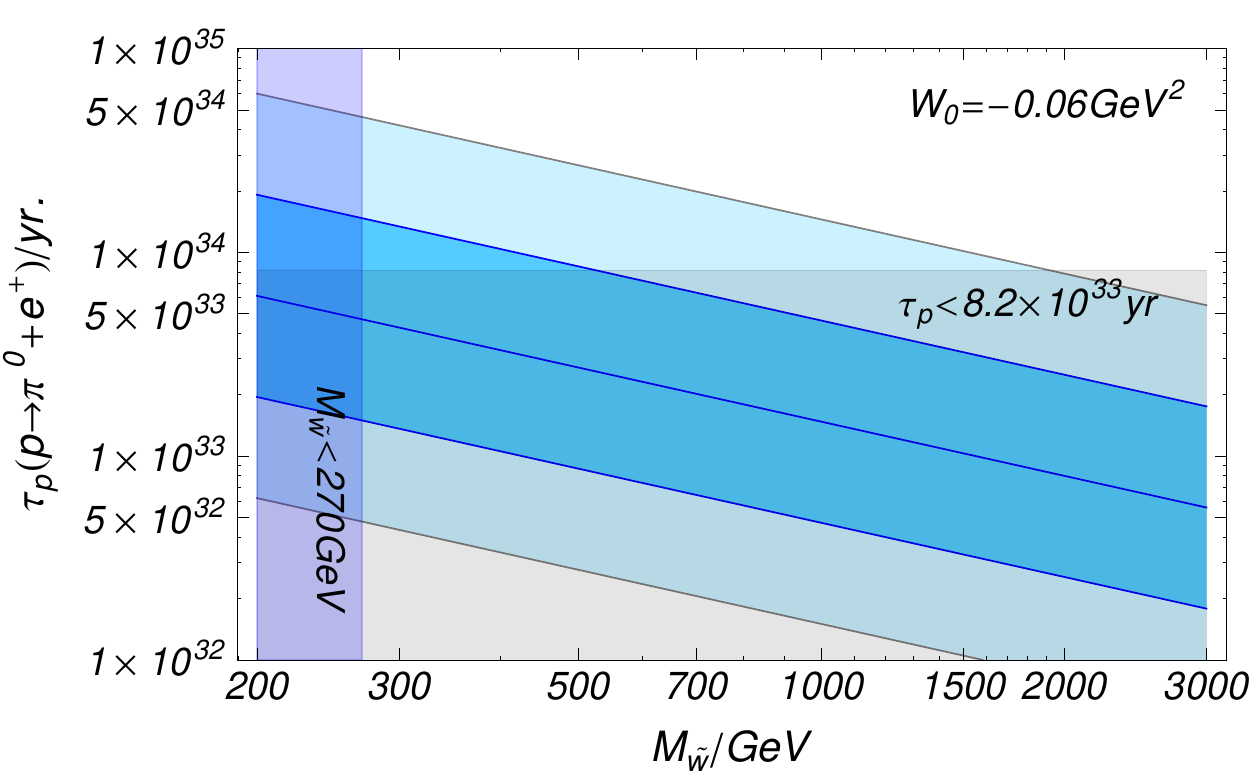}
 \end{minipage}
 \caption{\sl\small
 (Left) The mass range of the GUT gauge boson as a function of the 
 mass of the wino-like fermion. 
We have assumed $N_{\rm th}^{\rm (MAX)} = 5$ for  blue (dark) shaded region 
and $N_{\rm th}^{\rm (MAX)} = 10$ for light-blue (light) shaded region
in Eq.\,(\ref{eq:Mv}).
(Right) The proton lifetime via $p\to \pi^0+e^+$ assuming 
the GUT gauge boson mass in  Eq.\,(\ref{eq:Mv}).
Here, we have fixed the form factor to be $W_0 = -0.06$\,GeV$^2$ so that 
the exclusion limit is conservative.
The horizontally shaded region is excluded by the current lower limit $\tau_p \gtrsim 8.2\times 10^{33}$yr.
The vertically shaded region is excluded by the lower limit, $M_{\tilde w} \gtrsim 270$\,GeV.
 }
\label{fig:proton}
\end{figure}

Thinking along the same lines in section~\ref{sec:unification}, 
let us give an model independent estimation on $M_V$ in the following way.
First, we take $\mu = \Lambda = M_{\rm GUT}$ without loss of generality.
Then, let us define 
\begin{eqnarray}
{\mit \D} \alpha^{-1}_{ij} \equiv \frac{1}{2\pi} 
\left(
\beta_{i}^{(XY)} -
\beta_{j}^{(XY)} 
\right)
\log\frac{M_V}{M_{\rm GUT}}\ ,
\end{eqnarray}
which encodes how large GUT breaking threshold effects are required to match the three gauge 
coupling constants to the universal one at $\mu = M_{\rm GUT}$.
If ${\mit \D} \alpha^{-1}_{ij}$'s are too large, the nominal coupling unification 
in the adjoint extension of the SM is just an accidental one achieved by accidental cancellations 
between large GUT breaking contributions.
Thus, as in the track of the discussion in section\,\ref{sec:unification},
we require that ${\mit \D} \alpha^{-1}_{ij}$'s are not too large.
Concretely, we adopt the same criteria in Eq.\,(\ref{eq:NTH}), i.e.
\begin{eqnarray}
\label{eq:a13}
 |{\mit \D} \alpha_{13}| < \frac{N_{\rm th}^{\rm (MAX)}}{2\pi}\ .
\end{eqnarray}
Altogether, by substituting the $\b$'s in Eq.\,(\ref{eq:beta_GUT}), this requirement amounts to a
constraint on the GUT gauge boson mass,
\begin{eqnarray}
\label{eq:Mv}
 M_{V} = M_{\rm GUT}\times
  \exp\left[{\pm\left|\frac{N_{\rm th}^{\rm (MAX)}}{\beta_2^{(XY) }- \beta_1^{(XY)}} \right|}\right]
  =  M_{\rm GUT}\times\exp\left[{\pm\left|\frac{2N_{\rm th}^{\rm (MAX)}}{35} \right|}\right]\ ,
\end{eqnarray}
where we again take $N_{\rm th}^{\rm (MAX)}$ to be $5$ or $10$ as reference values.%
\footnote{
For given $M_{V}$  and $M_{\rm GUT}$, the largest $|{\mit \D}\alpha_{ij}|$ 
is ${\mit \D}\alpha_{12}$, and we may require
$|{{\mit \D} \alpha_{12}}| < {N_{\rm th}^{\rm (MAX)}}/{2\pi}$
instead of Eq.\,(\ref{eq:a13}), although it does not change
the following arguments significantly.
}
In the left panel of Fig.\,\ref{fig:proton}, we show the mass range of $M_V$ as a function
of $M_{\tilde w}$.
The figure shows that $M_V ={\cal O}(10^{15})$\,GeV in the whole mass range of the wino-like fermion.

%%%%%%%%%%%%%%%%%%%%%%%%%%%%%%%%%%%%%%%%%
%%%%%%%%%%%%%%%%%%%%%%%%%%%%%%%%%%%%%%%%%

The proton decay process, $p\to \pi^0+e^+$, proceeds through effective operators,
\begin{eqnarray}
{\cal L} = \frac{g_{\rm GUT}^2}{M_V^2}\left[ 
A_R\,(\bar{d}_R^{\dagger}\bar{u}_R^{\dagger})
(u_Le_L)
+
A_L(1+|V_{ud}|^2)
\,(u_Ld_L)(\bar{u}_R^{\dagger}\bar{e}_R^{\dagger}) + h.c.
\right]\ ,
\end{eqnarray}
 (see for example Ref.\,\cite{Ellis:1980jm}).
Here, $V_{ud}\simeq 0.974$  denotes the $ud$-component of the Cabibbo-Kobayashi-Masukawa matrix.
The coefficients  $A_{R,L}$ represent the renormalization factors of the 
above operators between the GUT scale to a lower energy scale.
At the renormalization scale $\m=2$\,GeV, the coefficients $A_{R, L}$ 
are given by,
\begin{eqnarray}
 A_{R,L} = A_{R,L}^{\rm SM}
 \times 
 \left( \frac{\alpha_2(M_{\tilde w})}{\alpha_{\rm GUT}} \right)^{\frac{9}{4}\left(
b_{2}^{-1}(\mu>M_{\tilde w}) -
b_{2}^{-1}(\mu<M_{\tilde w})
\right)
  }
  \times 
 \left( \frac{\alpha_3(M_{\tilde g})}{\alpha_{\rm GUT}} \right)^{2\left(
b_{3}^{-1}(\mu>M_{\tilde g}) -b_{3}^{-1}(\mu<M_{\tilde g})
\right)
}\ .
%&\simeq& A_{R,L}^{\rm SM}\times(1.0-1.2)\ ,
\end{eqnarray}
Here,  $A_{R,L}^{\rm SM}$ is the renormalization factor without having 
the adjoint fermions below the GUT scale which 
have been estimated to be $A_R^{\rm SM} \simeq 3$ and $A_L^{\rm SM} \simeq 3.2$
at $\m=2$\,GeV\,\cite{Ellis:1980jm}.
%The coefficients of the beta functions are given in Eq.\,(\ref{eq:betas}).
We find that $A_{R,L}$
is not very different from $A_{R,L}^{\rm SM}$, $A_{R,L}/A_{R,L}^{\rm SM} \simeq 1.0$--$1.2$,
for wide ranges of $M_{\tilde w}$ and  $M_{\tilde g}$.

Altogether, the resultant lifetime of the proton is given by,
\begin{eqnarray}
\tau(p\to \pi^0+e^+)\simeq 4.8\times 10^{33}\,{\rm yr}
\times
\left(\frac{A_{L,R}^{\rm SM}}{A_{L,R}}\right)^2
\left(\frac{1/40}{\a_G}\right)^2
\left(\frac{M_V}{10^{15.5}\,\rm GeV}\right)^4
\left(\frac{0.103\,{\rm GeV}^2}{|W_0|}\right)^2\ ,
\end{eqnarray}
where $W_0$ denotes the form factor of the proton decay operators 
between the proton and the pion states. 
In this analysis, we use $W_0 \simeq - 0.103$\,GeV$^2$ which is obtained
by lattice QCD simulations\,\cite{Aoki:2013yxa} with a total error about $30$--$40$\%.
In the right panel of Fig.\,\ref{fig:proton}, we show the predicted proton lifetime  assuming the GUT gauge boson mass
ranges in Eq.\,(\ref{eq:Mv}).
We also show the lower limit on the proton lifetime at 90\% confidence level
by the Super-Kamiokande with the total exposure $140$\,kton-yrs.
 $\tau_p > 8.2 \times 10^{33}$\,yr.~\cite{Nishino:2012ipa}. 
The figure shows that $M_{\tilde W}\gtrsim 550$\,GeV ($1.9$\,TeV) has been excluded due to a 
too short proton lifetime for $N_{\rm th}^{\rm (MAX)} = 5\, (10) $. 
The figure also shows that the whole mass range of the wino-like fermion 
can be surveyed by the Hyper-Kamiokande experiment
which is sensitive to the proton lifetime of $O(10^{35})$\,yr\,\cite{Abe:2011ts}.

Before closing our discussion, let us summarize other constraints on the wino-like fermions.
Due to the small mass difference between the charged and the neutral components of the 
wino-like fermion, the charged wino-like fermion has a rather long lifetime, and hence,
it leaves a disappearing  track once it is produced at the collider experiments.
By searching for the disappearing charged tracks, the ATLAS  collaboration has
put a stringent constraints on the mass of the wino-like fermion mass,
\begin{eqnarray}
M_{\tilde w} \gtrsim 270\,{\rm GeV}\ ,
\end{eqnarray}
with $20$\,fb$^{-1}$ data at 8\,TeV running~\cite{Aad:2013yna}.
At the 14\,TeV running, the limit is expected to be pushed up to $500$\,GeV with $100$\,fb$^{-1}$ data~\cite{Yamanaka}.
For more details on the future prospects of the searches for the wino-like fermion
at the collider experiments, see Ref.\,\cite{Cirelli:2014dsa}.

The mass of the wino-like fermion is also constrained from the indirect detection of dark matter 
using cosmic-rays.
Currently, the most robust limit comes from the continuum gamma-ray searches from dwarf spheroidal galaxies 
at the Fermi-LAT experiment which has excluded $M_{\tilde w} \lesssim 320$\,GeV and
$2.25$\,TeV$\lesssim M_{\tilde w} \lesssim 2.43$\,TeV 
at the 95\% confidence level using four-year data~\cite{Ackermann:2013yva}.%
\footnote{For uncertainties originating from the dark matter profile
and future prospects of the searches for the wino-like dark matter
via the gamma-rays from the dwarf spheroidal galaxies, see 
e.g.~\cite{Bhattacherjee:2014dya,Geringer-Sameth:2014qqa}. 
}
The searches for monochromatic gamma-rays
from the galactic center~\cite{Abramowski:2013ax} as well as the dwarf spheroidal galaxies~\cite{::2014ios}
by the H.E.S.S experiments also put constraints on the wino-like fermion mass.
Those constraints are, however, less stringent compared with the above constraints 
due to large uncertainties of the dark matter profile at the galaxy center 
(see e.g. Ref.\,\cite{Nesti:2013uwa}) 
and the small cross section into the monochromatic gamma-rays.%
\footnote{See Refs.\,\cite{Cohen:2013ama,Fan:2013faa} for related discussions.}
For the constraints from other cosmic-rays, see e.g. Ref.\,\cite{Hryczuk:2014hpa}.%
\footnote{As discussed in Ref.\,\cite{Ibe:2013jya} (see Refs.\,\cite{Shirai:2009fq} for earlier works),
the decaying wino-like dark matter in the MSSM via dimension six operators such as  $\tilde w \bar{e}_R\ell_L\ell_L$ 
can reproduce the excess of the positron fraction in the cosmic ray observed by PAMELA~\cite{Adriani:2008zr} and AMS-02 experiments~\cite{Aguilar:2013qda}.
In the case of the non-supersymmetric wino-like fermion, however,
the dimension six operator is dominated over by a dimension four operator $\tilde{w}\ell_L H^*$
which is at least  generated radiatively from the dimension six  operator, which 
is more constrained by the gamma-ray observations than the case with the 
dimension six operators (see e.g. Ref.\cite{Ibe:2013nka}).
}

The direct detection experiments of the wino-like fermion are, on the other hand, challenging for 
on-going experiments since it has no tree-level  $\tilde w-\tilde w-$Higgs nor $\tilde w-\tilde w-Z$ interactions.
As estimated in \cite{Hisano:2012wm}, 
however, loop suppressed interactions lead to a spin-independent nucleon cross section
of $\sigma_{\tilde w N} = {\cal O}(10^{-47})\,{\rm cm}^2$, with which it might be possible to test the model
by such as the LZ experiment \cite{Malling:2011va} and the DARWIN experiment \cite{Baudis:2012bc}
 for $M_{\tilde w} \lesssim 1$\,TeV.

\section{Conclusions}
In this paper, we revisited a small extension of the SM which achieve a precise coupling unification
while providing a good dark matter candidate, the wino-like fermion.
As we have discussed, the non-thermal production of the wino-like fermion from the gluino-like fermion
can dominate over the thermal contribution which allows the lighter wino-like dark matter 
than the minimal dark matter scenario.
With such a lighter wino-like dark matter, the proton lifetime is predicted above the 
current lower limit by the Super Kamiokande experiment. 
We also found that the most parameter space can be tested through the combination of the 
direct searches at the LHC experiments, the cosmic gamma-ray searches
and the search for the proton decay at the Hyper-Kamiokande experiment. 

As a final remark, let us comment on the possible warm component of the wino-like dark matter.
As we have discussed in section\,\ref{sec:DM}, most of the wino-like dark matter produced by the 
decay of the gluino-like fermion immediately blends into the thermal bath.
However, if the neutral wino-like fermion is produced with a very small momentum for $T_{\rm eff} = {\cal O}(100)$\,MeV
it cannot lose its momentum very efficiently, which may end up with a warm component.
Thus, a very small fraction of the wino-like dark matter can be a warm component of dark matter, 
which might leave some imprints on the small scale structure\,\cite{Ibe:2012hr}.

\section*{Acknowledgements}
We thank Shigeki Mastumoto for sharing his note on the Sommerfeld enhancement.
We also thank Keisuke Harigaya and Shigeki Mastumoto for useful discussion on the non-thermal production
of the wino-like fermion by the decay of the very heavy gluino-like fermion.
This work is supported by Grant- in-Aid for Scientific research from the Ministry of Education, Science, Sports, and Culture (MEXT), Japan, No. 24740151 and 25105011 (M.I.), from the Japan Society for the Pro- motion of Science (JSPS), No. 26287039 (M.I.), the World Premier International Research Center Initiative (WPI Initiative), MEXT, Japan (M.I.).
The authors are grateful to Kavli-IPMU

\appendix
\section{Fate of the high energetic quarks}
\label{sec:thermalization}
In this appendix, let us discuss how the produced quarks by the decay of the gluino-like fermion 
form the thermal bath.
When the universe is dominated by radiation, the high energy quarks emitted by the decay of the 
gluino-like fermion lose their energy very quickly and is resolved into radiation immediately.
Even after the gluino-like fermion dominates the universe, the radiation energy is dominated 
by the original radiation for $T > T_*$ in Eq.\,(\ref{eq:Tstar}).
Thus, the produced quark is again resolved into the radiation quickly for $T>T_*$.

Once the temperature becomes lower than $T_*$, on the other hand, the radiation energy
is dominated by the one from the decay of the gluino-like fermion.
Since each quark has much higher energy than the temperature, the number density
of the quarks are much smaller than the one of the radiation in the thermal equilibrium.
Thus, in this period, there should be efficient inelastic interactions which change 
the number of the particles so that the produced quarks form the thermal radiation.

In our scenario, the thermal bath exists before the decay of the gluino-like fermion,
and hence, the thermalization proceeds via inelastic interactions between
the injected quarks and the pre-existing thermal bath.
The energy loss rate in the thermalization process is given by~\cite{Harigaya:2013vwa} (see also \cite{Kurkela:2011ti}),
\begin{eqnarray}
 \frac{d E}{dt} \sim \alpha^2 T^2 \sqrt{\frac{E}{T}}\ ,
\end{eqnarray}
where $T$ is the temperature of the pre-existing thermal bath. 
Thus, a typical time-scale for the high-energetic quarks ends up with the thermal radiation is given by,
\begin{eqnarray}
  t_{\rm split} \sim (\alpha^2 T)^{-1} \sqrt{\frac{M_{\tilde g}}{T}}\ .
\end{eqnarray}
The Hubble scale scales by $a^{-3/2}$ while $t_{\rm split}^{-1}$ scales by $a^{-3/2}$
for $T>T_*$, and by $a^{-9/16}$ for $T < T_*$.
Thus, the thermalization process is always effective if $t_{\rm split}^{-1}/H\gg 1$
at $T \simeq T_{\rm dom}$.
By remembering Eq.\,(\ref{eq:Tdom}), we immediately find that $t_{\rm split}^{-1}/H_{\rm dom} \gg 1$,
and hence, the high-energetic quarks are thermalized immediately.

\end{document}